\documentclass[aps, rmp, twocolumn]{revtex4-2}
\usepackage{amsmath}
\usepackage{amssymb}

\usepackage{color}

\usepackage{graphicx}

\usepackage{hyperref}

\usepackage{multirow}
\usepackage{array}

\allowdisplaybreaks

\begin{document}

\title{Localization in Quantum Field Theory}

\author{Riccardo Falcone}
\affiliation{Department of Physics, University of Sapienza, Piazzale Aldo Moro 5, 00185 Rome, Italy}

\author{Claudio Conti}
\affiliation{Department of Physics, University of Sapienza, Piazzale Aldo Moro 5, 00185 Rome, Italy}

\begin{abstract}
We review the issue of localization in quantum field theory and detail the nonrelativistic limit. Three distinct localization schemes are examined: the Newton-Wigner, the algebraic quantum field theory, and the modal scheme. Among these, the algebraic quantum field theory provides a fundamental concept of localization, rooted in its axiomatic formulation. In contrast, the Newton-Wigner scheme draws inspiration from the Born interpretation, applying mainly to the nonrelativistic regime. The modal scheme, relying on the representation of single particles as positive frequency modes of the Klein-Gordon equation, is found to be incompatible with the algebraic quantum field theory localization.

This review delves into the distinctive features of each scheme, offering a comparative analysis. A specific focus is placed on the property of independence between state preparations and observable measurements in spacelike separated regions. Notably, the notion of localization in algebraic quantum field theory violates this independence due to the Reeh-Schlieder theorem. Drawing parallels with the quantum teleportation protocol, it is argued that causality remains unviolated. Additionally, we consider the nonrelativistic limit of quantum field theory, revealing the emergence of the Born scheme as the fundamental concept of localization. Consequently, the nonlocality associated with the Reeh-Schlieder theorem is shown to be suppressed under nonrelativistic conditions.
\end{abstract}
\maketitle

\tableofcontents

\section{Introduction}

In the NonRelativistic Quantum Mechanics (NRQM), the notion of localization is notoriously given in terms of wave functions and position operator and follows Born's interpretation of quantum mechanics. States are localized in the support of their wave functions, whereas second-quantized observables are localized in $\vec{x}$ if they are generated by creators and annihilators of particles in $\vec{x}$. Also, states are orthogonal if the supports of their wave functions are disjoint and, hence, if they are localized in different regions.

This notion of localization was then extended to QFT by \citet{RevModPhys.21.400} and by \citet{fulling_1989}. The so-called Newton-Wigner localization is based on the orthogonality condition between states in disjoint regions and other natural requirements that make it conceptually equivalent to the Born scheme. At variance with Born, however, \citet{RevModPhys.21.400} worked in the context of relativistic theories.

Notably, the Hegerfeldt theorem \cite{PhysRevD.10.3320} forbids any notion of localization that assumes causal propagation of wave functions and orthogonality condition between states in disjoint regions of space. The resulting superluminal propagation of wave functions in the Newton-Wigner scheme is unsatisfactory as it violates the relativistic postulate of causality in QFT. This led to the idea that the Newton-Wigner scheme is not suited for a genuine description of local phenomena in QFT. 

A fundamental notion of localization is, instead, provided by the Algebraic Quantum Field Theory (AQFT) formalism \cite{10.1063/1.1704187, haag1992local, Brunetti:2015vmh, 10.1007/978-3-030-38941-3_1, Halvorson:2006wj}. The algebraic approach to QFT is based on the definition of local algebras by means of quantum fields in spacetime points. This gives a natural definition of local observables and local preparation of states. 

At variance with the Newton-Wigner scheme, the AQFT framework provides a genuine notion of localization and it faithfully describes local experiments in isolated laboratories. The main argument is that the instantaneous propagation of Newton-Wigner wave functions is in conflict with relativity. Conversely, in AQFT, the microcausality condition appears as an axiom of the theory and is expressed by the commutativity of quantum fields in spacelike separated points.

Unfortunately, the microcausality condition does not guarantee the independence between the preparation of states and the measurement of observables in spacelike separated regions. The Reeh-Schlieder theorem \cite{Reeh:1961ujh} demonstrates that states that are localized with respect to the AQFT scheme are not necessarily strictly localized \cite{10.1063/1.1703731, 10.1063/1.1703925}. Explicitly, this means that the outcome of measurements in a region $\mathcal{O}_\text{B}$ may depend on the local preparation of states in $\mathcal{O}_\text{A}$ even if $\mathcal{O}_\text{A}$ and $\mathcal{O}_\text{B}$ are causally disconnected.  However, it has been argued that such a nonlocal effect does not violate causality since it only comes from selective nonunitary preparations of states \cite{Redhead1995-REDMAA-2, CLIFTON20011, VALENTE2014147, RevModPhys.90.045003}.

At variance with the AQFT, the Newton-Wigner scheme is not affected by the Reeh-Schlieder nonlocality. In particular, any state localized with respect to the Newton-Wigner scheme is also strictly localized and does not affect measurements in any other disjoint space regions. In this case, we say that the strict localization property is always satisfied.

Also, as a consequence of a corollary of the Reeh-Schlieder theorem \cite{Redhead1995-REDMAA-2, RevModPhys.90.045003}, no local creator and annihilator operator can be defined in the AQFT formalism. This means that, contrary to the Newton-Wigner scheme, the AQFT scheme is not characterized by local Fock spaces and local vacua.

All the incompatibilities between the two localization schemes disappear in the nonrelativistic limit. In particular, it has been proven that any operator that is localized in $\vec{x}$ with respect to the Newton-Wigner scheme approximates to an operator localized in $\vec{x}$ with respect to the AQFT scheme \cite{PhysRevD.90.065032, Papageorgiou_2019}. This result is in agreement with the fact that the Born-Newton-Wigner and the AQFT schemes are suited for the description of phenomena in, respectively, the nonrelativistic and the relativistic regime.

Owing to the convergence to the AQFT, the Newton-Wigner scheme becomes valid in the nonrelativistic regime. Furthermore, the Reeh-Schlieder nonlocal effect is suppressed and any state locally prepared in a space region $\mathcal{V}_1$ is also strictly localized in $\mathcal{V}_1$, even if the preparation is a selective nonunitary operation.

In addition to the Newton-Wigner and the AQFT scheme, we study the modal localization scheme, which is based on the representation of single particles as positive frequency modes of the Klein-Gordon equation \cite{Wald:1995yp}. The fact that states are localized in the support of the corresponding modes is generally inaccurate, since a genuine notion of localization is only given by the AQFT formalism which is incompatible with the modal scheme in the relativistic regime. However, in the nonrelativistic limit, the two localization schemes converge.

In our previous works \cite{PhysRevD.107.045012, PhysRevD.107.085016, PhysRevA.108.022807}, we assumed that nonrelativistic states are localized in the support of their modal wave functions. Here, we find a justification for such a claim by demonstrating the convergence between the modal and the AQFT scheme.

This manuscript is intended to be an introductory review on the problem of localization in QFT. We consider a Minkowski spacetime $\mathcal{M}$ represented by the coordinates $x^\mu = (t,\vec{x})$ and characterized by the flat metric $\eta^{\mu\nu} = \text{diag}(-c^{-2},1,1,1)$. We focus on a real scalar field $\hat{\phi}(x^\mu)$ since the only important elements of the theory are captured by quantum fields without internal degrees of freedom. The corresponding quantum states are elements of the Minkowski-Fock space $\mathcal{H}_\text{M}$ constructed from the creators of the Klein-Gordon particles with defined momenta $\hat{a}^\dagger(\vec{k})$ and from the Minkowski vacuum $| 0_\text{M} \rangle$, which is defined as the state annihilated by $\hat{a}(\vec{k})$, i.e., $\hat{a}(\vec{k}) | 0_\text{M} \rangle = 0$ for any $\vec{k} \in \mathbb{R}^3$. The algebra of operators acting on $\mathcal{H}_\text{M}$ will be noted as $\mathfrak{A}_\text{M}$.

The paper is organized as follows. In Secs.~\ref{NewtonWigner_localization_scheme}, \ref{AQFT_localization_scheme} and \ref{Modal_localization_scheme} we define the Newton-Wigner, the AQFT and the modal schemes, respectively. Their features are then shown and compared to each other in Sec.~\ref{Comparison_between_localization_schemes}. In Sec.~\ref{Localization_in_NRQM}, we study the localization in NRQM; in particular, we define the Born scheme and we show the convergence of all localization schemes in the nonrelativistic limit. Conclusions are drawn in Sec.~\ref{Localization_in_Quantum_Field_Theory_Conclusions}.

\section{Newton-Wigner scheme}\label{NewtonWigner_localization_scheme}

\citet{RevModPhys.21.400} addressed the problem of localization of particles in Relativistic Quantum Mechanics (RQM) by deriving the position observable and its eigenstates from first principles. They showed that the definition of localization is uniquely determined by some natural requirements. They assumed the following general theoretic postulates on the basis of which a particle can be considered localized at time $t=0$ in $\vec{x}$: (i) the superposition of localized states is localized as well; (ii) the set of localized states in $\vec{x}$ is invariant under rotations and time and space reflections with $\vec{x}$ as a fixed point; (iii) states localized in different spatial positions $\vec{x} \neq \vec{x}'$ are orthogonal; (iv) some regularity conditions of mathematical good behavior. From these assumptions, \citet{RevModPhys.21.400} derived the definition of a unique position operator $\hat{\vec{x}}_\text{NW}$ and localized states $|  \vec{x}_\text{NW} \rangle$. The operator was then second quantized by \citet{fulling_1989}, who reformulated the theory in the context of QFT.

\citet{RevModPhys.21.400} started from the representation of the spinless elementary particles (i.e., Klein-Gordon single particles) via irreducible representation of the Poincare group (i.e., energy, momentum and angular momentum). Then, they studied the case of particles with spin and finite mass. The uniqueness of the position operator satisfying the natural transformation conditions in RQM with arbitrary spin was later discussed by \citet{Weidlich_Mitra} and led to the same conclusions as \citet{RevModPhys.21.400}.

The Newton-Wigner scheme in QFT predicts a phenomenon of superluminal spreading \cite{PhysRev.139.B963} that is in contrast with the relativistic notion of causality. This is a consequence of the Hegerfeldt theorem \cite{PhysRevD.10.3320, PhysRevD.22.377}, whose only hypotheses are the positivity of the energy of relativistic particles and the orthogonality condition of states localized in disjoint regions. Due to the violation of causality, the Newton-Wigner scheme is not regarded as fundamental in nature. Conversely, in Sec.~\ref{AQFT_localization_scheme}, we will see that the notion of locality provided by the AQFT does not lead to superluminal signaling and can be regarded as a genuine localization scheme in QFT.

The present section is organized as follows. In Sec.~\ref{NewtonWigner_scheme_in_RQM}, we briefly show the results of Newton and Wigner's work in the context of RQM. Then, in Sec.~\ref{NewtonWigner_scheme_in_QFT} we consider the second-quantized version of the position operator to define the Newton-Wigner scheme in QFT. Lastly, in Sec.~\ref{Hegerfeldt_theorem}, we review the literature about the Hegerfeldt theorem and its consequences on the problem of localization.

\subsection{Newton-Wigner scheme in RQM}\label{NewtonWigner_scheme_in_RQM}

Here, we give the definition of Newton-Wigner position operator and wave functions in the context of RQM.

States with defined momentum $| \vec{k} \rangle$ are defined as eigenstates of the momentum operator $\hat{\vec{k}}$, satisfying the following orthonormalization property
\begin{equation}\label{relativistic_orthonormalization}
\langle \vec{k} | \vec{k}' \rangle = \delta^3(\vec{k}-\vec{k}').
\end{equation}
Starting from the definition of $| \vec{k} \rangle$, \citet{RevModPhys.21.400} derived the unique state satisfying conditions (i)-(iv) as 
\begin{equation}\label{x_state}
|  \vec{x}_\text{NW} \rangle = \int_{\mathbb{R}^3} d^3 k \frac{e^{-i \vec{k} \cdot \vec{x}}}{\sqrt{(2 \pi)^3}} | \vec{k} \rangle.
\end{equation}

Equation (\ref{x_state}) describes the state localized in $\vec{x}$ at time $t=0$ according to the Newton-Wigner scheme in RQM. This provides the definition of the position operator
\begin{equation}\label{x_NW}
\hat{\vec{x}}_\text{NW} = \int_{\mathbb{R}^3} d^3 x \vec{x} | \vec{x}_\text{NW} \rangle \langle \vec{x}_\text{NW} |,
\end{equation}
whose eigenstates are $|  \vec{x}_\text{NW} \rangle$ with eigenvalues $\vec{x}$. Also, for any state $| \psi \rangle$, \citet{RevModPhys.21.400} defined the wave function in position space as
\begin{equation}\label{psi_NW}
\psi_\text{NW}(t,\vec{x}) = \langle \vec{x}_\text{NW} | \psi(t) \rangle = \int_{\mathbb{R}^3} d^3 k \frac{ e^{-i \omega(\vec{k}) t+i \vec{k} \cdot \vec{x}}}{\sqrt{(2 \pi)^3}} \tilde{\psi}(\vec{k}),
\end{equation}
with
\begin{equation} \label{omega_k}
\omega(\vec{k}) = \sqrt{\left(\frac{mc^2}{\hbar}\right)^2 + c^2 |\vec{k}|^2}
\end{equation}
as the frequency of $\vec{k}$ and with $\tilde{\psi}(\vec{k})$ as the wave function in momentum space, defined as
\begin{equation}
\tilde{\psi}(\vec{k}) = \langle \vec{k} | \psi \rangle.
\end{equation}
Here, $m$ is the mass of the particle. The inner product between states can be written in terms of their wave function in position space as
\begin{equation}\label{psi_NW_product}
\langle \psi | \psi' \rangle = \int_{\mathbb{R}^3} d^3 x \psi_\text{NW}^*(t,\vec{x}) \psi'_\text{NW}(t,\vec{x}),
\end{equation}
which is the familiar $L^2(\mathbb{R}^3)$ scalar product.

Notice that in NRQM, wave functions in momentum space are related to wave functions in position space by means of the Fourier transform. The same occurs in RQM between the wave functions $\tilde{\psi}(\vec{k})$ and $\psi_\text{NW}(0,\vec{x})$ [Eq.~(\ref{psi_NW})]. This analogy leads to the equivalence between the Newton-Wigner and the Born localization schemes, which will be detailed in Sec.~\ref{Born_localization_is_NewtonWigner_localization}.

\citet{RevModPhys.21.400} already pointed out in their original work that the position operator $\hat{\vec{x}}_\text{NW}$ is not relativistically covariant. For any Lorentz boost $\Lambda_{\vec{v}}: (t,\vec{x}) \mapsto (t',\vec{x}')$, the state that is localized in (say) $\vec{x}=0$ at $t=0$ is not localized in $\vec{x}' = 0$ at $t'=0$. Hence, two inertial observers do not share the same notion of localization. This is an important argument against the Newton-Wigner localization program, since Lorentz transformed frames are physically equivalent in relativistic theories.

Furthermore, the Newton-Wigner localization is found not to be preserved in time. Specifically, a particle localized in a bounded region at $t=0$ will develop infinite tails at immediately later times $t \neq 0$, exceeding the light cone of the initial region \cite{PhysRev.139.B963}. The phenomenon of superluminal spreading of the wave functions was then proved to occur for a more general class of localization schemes. The only condition is a nonconstant Hamiltonian that is a semibounded function of the particle momentum \cite{PhysRevD.22.377}. This model-independent result goes under the name of Hegerfeldt theorem and will be discussed in Sec.~\ref{Hegerfeldt_theorem}.

The non-covariant behavior of the position operator $\hat{\vec{x}}_\text{NW}$ and the acausal spreading of the wave functions make the Newton-Wigner localization unsatisfactory for a fully relativistic theory. The solution to this problem will be found by noticing that the operator $\hat{\vec{x}}_\text{NW}$ does not entail any fundamental notion of locality; conversely, it is a mathematical artifice that comes from the nonrelativistic theory. Only in the nonrelativistic limit of RQM, the Newton-Wigner scheme obtains a genuine notion of locality. This result will be shown in Sec.~\ref{Comparison_with_the_relativistic_theory}.

\subsection{Newton-Wigner scheme in QFT}\label{NewtonWigner_scheme_in_QFT}

In Sec.~\ref{NewtonWigner_scheme_in_RQM}, we worked in the context of RQM and used the definition of first-quantized Newton-Wigner position operator $\hat{\vec{x}}_\text{NW}$ to define localized states and wave functions in position space. Here, we apply these results to the framework of QFT. In particular, we define localized states and observables by means of a second-quantized version of $\hat{\vec{x}}_\text{NW}$. The method is based on the natural embedding of RQM in QFT as the theory of single particle states of the corresponding quantum fields.

For any scalar field $\hat{\phi}(t,\vec{x})$, the corresponding single particle state with defined momentum $\vec{k}$ is defined as
\begin{equation}\label{single_particle_field_embedding}
| \vec{k} \rangle = \hat{a}^\dagger(\vec{k}) | 0_\text{M} \rangle ,
\end{equation}
where $\hat{a}^\dagger(\vec{k})$ is a creator operator satisfying the canonical commutation identities
\begin{align}\label{Minkowski_canonical_commutation}
& [\hat{a}(\vec{k}),\hat{a}^\dagger(\vec{k}')]  = \delta^3(\vec{k}-\vec{k}'),
& [\hat{a}(\vec{k}),\hat{a}(\vec{k}')] = 0.
\end{align}
Single particle states with defined position are defined by Eq.~(\ref{x_state}). Owing to Eq.~(\ref{single_particle_field_embedding}), Eq.~(\ref{x_state}) is equivalent to
\begin{equation}
| \vec{x}_\text{NW} \rangle = \hat{a}_\text{NW}^\dagger(\vec{x}) | 0_\text{M} \rangle,
\end{equation}
where
\begin{equation}\label{a_NW}
\hat{a}_\text{NW}(\vec{x}) = \int_{\mathbb{R}^3} d^3 k \frac{e^{i \vec{k} \cdot \vec{x}}}{\sqrt{(2 \pi)^3 }} \hat{a}(\vec{k})
\end{equation}
is the inverse of the Fourier transform of the annihilation operator $\hat{a}(\vec{k})$.

Notice that there is a one-to-one mapping between the operators $\hat{a}_\text{NW}(\vec{x})$ and $\hat{a}(\vec{k})$ for varying $\vec{x}$ and $\vec{k}$. Hence, the entire Minkowski-Fock algebra $\mathfrak{A}_\text{M}$ is generated by $\hat{a}_\text{NW}(\vec{x})$ with varying $\vec{x}$, in the sense that any operator acting on the Minkowski-Fock space $\mathcal{H}_\text{M}$ can be written as a linear combination of products of $\hat{a}_\text{NW}(\vec{x})$ and $\hat{a}_\text{NW}^\dagger(\vec{x})$ operators.

Hereafter, the algebra generated by $\hat{a}_\text{NW}(\vec{x})$ with fixed $\vec{x}$ is denoted as $\mathfrak{A}_\text{M}^\text{NW}(\vec{x})$. We say that $\mathfrak{A}_\text{M}^\text{NW}(\vec{x})$ is a local algebra with respect to the Newton-Wigner scheme. Any element of $\mathfrak{A}_\text{M}^\text{NW}(\vec{x})$ is an operator that is localized in $\vec{x}$. Conversely, any local state $| \psi \rangle$ is the result of local operations on the vacuum background $| 0_\text{M} \rangle$. Hence, $| \psi \rangle$ is said to be localized in $\vec{x}$ if there is a local operator $\hat{O} \in \mathfrak{A}_\text{M}^\text{NW}(\vec{x})$ such that $| \psi \rangle = \hat{O} | 0_\text{M} \rangle$.

The definition of localized states and observables can also be generalized to extended regions. For any region $\mathcal{V} \subset \mathbb{R}^3$, we define $\mathfrak{A}_\text{M}^\text{NW}(\mathcal{V})$ as the local algebra in $\mathcal{V}$ generated by the operators $\hat{a}_\text{NW}(\vec{x})$ with $\vec{x} \in \mathcal{V}$. We say that the operator $\hat{O}$ and the state $| \psi \rangle = \hat{O} | 0_\text{M} \rangle$ are localized in $\mathcal{V}$ if $\hat{O}$ is an element of $\mathfrak{A}_\text{M}^\text{NW}(\mathcal{V})$.

The definitions of local operators and states provided here come from a second-quantized generalization of $\hat{\vec{x}}_\text{NW}$. Notice that, by embedding the relativistic theory of single particles, the Newton-Wigner scheme in QFT inherits all the issues concerning the localization of states described by Sec.~\ref{NewtonWigner_scheme_in_RQM}. This includes the instantaneous propagation of localized states and the consequent violation of causality, which will be discussed in the next subsection.

\subsection{Hegerfeldt theorem}\label{Hegerfeldt_theorem}

\citet{PhysRevD.10.3320} showed that the phenomenon of instantaneous spreading for a relativistic particle does not occur only in the Newton-Wigner scenario. An alternative proof was later provided by \citet{PhysRevD.16.315}. \citet{PhysRevD.22.377} recognized that relativity is not needed to prove the results, while positivity of the energy and translation invariance suffice to give the instantaneous spreading. Then, \citet{10.1007/BFb0106784} recognized that translation invariance is also not needed and, hence, the role of positivity of energy appears to be crucial in the instantaneous spreading of the wave function. However, when translation invariance is not considered, the localized particle either develops infinite tails immediately after or stays in its support indefinitely.

\citet{PhysRevD.22.377} showed that any particle confined in a bounded region can be found in spacelike separated regions at later times if the Hamiltonian is a nonconstant semibounded function of the momentum and translation invariant. Under stronger assumptions, the spreading of the wave function is over all of space. The conditions considered by \citet{PhysRevD.22.377} are met in RQM and QFT, where the energy of particles $\omega(\vec{k})$ is a function of the momentum $\vec{k}$ and is always positive.

The generality of the results is given by the fact that no specific definition of localization has been considered. To prove that no state can be localized in a finite region for a finite time interval, \citet{PhysRevD.22.377} only assumed that states localized in disjoint regions are orthogonal to each other. Also, to show that the spreading is over all of space, the authors assumes the existence of a positive operator $\hat{N}(\mathcal{V})$ for any space region $\mathcal{V} \subset \mathbb{R}^3$, such that $\langle \psi | \hat{N}(\mathcal{V}) | \psi \rangle \in [0,1]$ is the probability of finding the particle in $\mathcal{V}$. For instance, in the case of Newton-Wigner localization, $\hat{N}(\mathcal{V})$ is defined as
\begin{equation}\label{N_NewtonWigner}
\hat{N}(\mathcal{V}) = \frac{1}{|\mathcal{V}|} \int_\mathcal{V} d^3 x \hat{a}_\text{NW}^\dagger(\vec{x}) \hat{a}_\text{NW}(\vec{x}),
\end{equation}
where $|\mathcal{V}|$ is the volume of $\mathcal{V}$.

The apparent contradiction with the causal nature of the Klein-Gordon equation (or any other hyperbolic equation satisfying finite propagation speed, e.g., Maxwell equation, Dirac equation) was argued by \citet{Afanasev:1996nm}. Given any positive frequency solution of the Klein-Gordon equation $\psi(t,\vec{x})$ such that $\psi(0,\vec{x}) = 0$ for any $\vec{x}$ outside $\mathcal{V}$, then one finds that $\psi(t,\vec{x}) = 0$ in any region spacelike distant from $\mathcal{V}$ only if $\partial_t \psi(t,\vec{x}) |_{t=0} = 0$, which does not occur for positive frequency solutions. This means that the localization in a finite region for a finite time is only possible for superpositions of positive and negative frequency solutions of the Klein-Gordon equation, which are excluded by the hypotheses of the Hegerfeldt theorem.

\section{AQFT scheme}\label{AQFT_localization_scheme}

In Sec.~\ref{NewtonWigner_localization_scheme} we reviewed the Newton-Wigner approach to the problem of localization in QFT. We remarked that the assumption made by \citet{RevModPhys.21.400} are included in the hypotheses of the Hegerfeldt theorem. The results of the theorem are incompatible with the causality principle, as they imply a superluminal propagation of the localization condition. The paradox can be resolved by noticing that, in QFT, the spacetime coordinates $x^\mu$ appear as variables of the fields $\hat{\phi}(x^\mu)$ and the causality condition is defined via commutativity of spacelike separated fields.

In the framework of AQFT \cite{10.1063/1.1704187, haag1992local, Brunetti:2015vmh, 10.1007/978-3-030-38941-3_1, Halvorson:2006wj}, any spacetime event $\mathcal{E} \in \mathcal{M}$ is provided with the local algebra $\mathfrak{A}(\mathcal{E})$, which is generated by the operator $\hat{\phi}(x^\mu)$ with $x^\mu$ as the Minkowski coordinate representing $\mathcal{E}$.\footnote{This statement is mathematically imprecise. In AQFT, local algebras are rigorously defined with respect to extended spacetime regions $\mathcal{O}$. The algebra $\mathfrak{A}(\mathcal{E})$ has to be considered in the limiting case of spacetime regions approximated by point-like events, i.e., $\mathcal{O} \rightarrow \mathcal{E}$.} Any element of $\mathfrak{A}(\mathcal{E})$ is a linear combination of powers of $\hat{\phi}(x^\mu)$. More generally, for any spacetime region $\mathcal{O} \subset \mathcal{M}$, the local algebra $\mathfrak{A}(\mathcal{O})$ is generated by the field $\hat{\phi}(x^\mu)$ smeared out with test functions that are supported in the Minkowski coordinate region $\mathcal{O}_\text{M} \subset \mathbb{R}^4$ representing $\mathcal{O}$. In other words, the operator $\hat{O}$ is an element of $\mathfrak{A}(\mathcal{O})$ if there are some functions $f_n(x^\mu_1, \dots, x^\mu_n)$ such that
\begin{align}\label{phi_f}
\hat{O} = & \sum_n \int_{\mathcal{O}_\text{M}}d^4 x_1 \cdots \int_{\mathcal{O}_\text{M}}d^4 x_n f_n(x^\mu_1, \dots, x^\mu_n) \nonumber \\
& \times \hat{\phi}(x^\mu_1) \cdots \hat{\phi}(x^\mu_n).
\end{align}

The operator $\hat{O}$ is said to be localized in $\mathcal{O}$ with respect to the AQFT scheme if $\hat{O}$ is an element of $\mathfrak{A}(\mathcal{O})$. We also define localized states by means of the notion of preparation over the vacuum $| 0_\text{M} \rangle$. The state $| \psi \rangle$ is said to be localized in $\mathcal{O}$ if it is the result of local operations on $| 0_\text{M} \rangle$. Explicitly, this means that $| \psi \rangle = \hat{O} | 0_\text{M} \rangle$, with $\hat{O} \in \mathfrak{A}(\mathcal{O})$.

The causality condition states that if $\mathcal{O}_\text{A}$ and $\mathcal{O}_\text{B}$ are spacelike separated regions, the corresponding local algebras $\mathfrak{A}(\mathcal{O}_\text{A})$ and $\mathfrak{A}(\mathcal{O}_\text{B})$ mutually commute. The commutativity of $\mathfrak{A}(\mathcal{O}_\text{A})$ and $\mathfrak{A}(\mathcal{O}_\text{B})$ imposes statistical independence of measurements in the spacelike separated regions $\mathcal{O}_\text{A}$ and $\mathcal{O}_\text{B}$, in the sense that measurements in $\mathcal{O}_\text{A}$ and $\mathcal{O}_\text{B}$ do not influence each other. This is known as the microcausality axiom of AQFT.

The microcausality condition is satisfied by the algebra of Klein-Gordon field $\hat{\phi}(x^\mu)$, due to canonical commutation relation
\begin{equation}\label{phi_commutation}
\left[ \hat{\phi}(t,\vec{x}), \hat{\phi}(t',\vec{x}') \right] = i \hbar \Delta_\text{KG}(t-t', \vec{x} - \vec{x}'),
\end{equation}
with
\begin{align}\label{PauliJordan_function}
\Delta_\text{KG}(t,\vec{x}) = &  - \frac{i}{(2 \pi)^3} \int_{\mathbb{R}^3} \frac{d^3 k}{2 \omega(\vec{k})} \nonumber \\
& \times \left[ e^{-i\omega(\vec{k})t + i\vec{k} \cdot \vec{x}} - e^{i\omega(\vec{k})t - i\vec{k} \cdot \vec{x}} \right]
\end{align}
as the Pauli–Jordan function. From Eqs.~(\ref{phi_commutation}) and (\ref{PauliJordan_function}), we find that $[ \hat{\phi}(t,\vec{x}), \hat{\phi}(t',\vec{x}') ] = 0$ if $(t,\vec{x})$ and $(t',\vec{x}')$ are spacelike separated. 

The microcausality axiom ensures independence of measurements in spacelike separated regions $\mathcal{O}_\text{A}$ and $\mathcal{O}_\text{B}$. However, measurements are not the only types of operations that can be carried out in local experiments. For instance, one may consider local preparations of states in $\mathcal{O}_\text{A}$ and test their influence on $\mathcal{O}_\text{B}$. The independence between preparations in $\mathcal{O}_\text{A}$ and measurements in $\mathcal{O}_\text{B}$ are not guaranteed by the microcausality axiom.

The localization program in AQFT is crucially affected by the Reeh-Schlieder theorem \cite{Reeh:1961ujh}, which predicts the presence of nonlocal quantum correlations in the vacuum $| 0_\text{M} \rangle$ \cite{haag1992local, Redhead1995-REDMAA-2, PhysRevA.58.135}. One of the consequences of the Reeh-Schlieder theorem is that measurements made in $\mathcal{O}_\text{B}$ are able to distinguish the vacuum $| 0_\text{M} \rangle$ from some states $| \psi \rangle$ localized in $\mathcal{O}_\text{A}$, even if $\mathcal{O}_\text{B}$ is spacelike separated from $\mathcal{O}_\text{A}$. Notwithstanding this apparent incompatibility with causality, it can be shown that the Reeh-Schlieder nonlocality cannot be used for superluminal signaling \cite{Redhead1995-REDMAA-2, CLIFTON20011, VALENTE2014147, RevModPhys.90.045003}.

The explicit hypotheses and statement of the Reeh-Schlieder theorem will be given in Sec.~\ref{ReehSchlieder_theorem}. Conversely, the solution to the apparent violation of causality will be discussed in Sec.~\ref{solving_the_paradox}.

\subsection{Reeh-Schlieder theorem}\label{ReehSchlieder_theorem}

In this subsection, we show the hypotheses and statement of the Reeh-Schlieder theorem. We discuss the dependency of spacelike separated operations and the nonlocality of number operators as consequences of the theorem.

The axioms of AQFT in flat spacetime are
\begin{enumerate}
\item Microcausality: operators in spacelike separated regions commute, i.e., $[ \hat{O}_\text{A}, \hat{O}_\text{B} ] = 0$ when $\hat{O}_\text{A} \in \mathfrak{A}(\mathcal{O}_\text{A})$ and $\hat{O}_\text{B} \in \mathfrak{A}(\mathcal{O}_\text{B})$ for any couple of spacelike separated regions $\mathcal{O}_\text{A}$ and $\mathcal{O}_\text{B}$;
\item Isotony: any observable in $\mathcal{O}$ can also be measured in a larger region $\mathcal{O}'$, hence, $\mathfrak{A}(\mathcal{O}) \subset \mathfrak{A}(\mathcal{O}')$ if $\mathcal{O} \subset \mathcal{O}'$;
\item Relativistic covariance: each Poincar\'e transformation $\rho$ is provided with a unitary representation $\hat{U}(\rho)$ such that $\hat{U}(\rho) \mathfrak{A}(\mathcal{O}) \hat{U}^\dagger(\rho) = \mathfrak{A}(\rho(\mathcal{O}))$ with the vacuum $|0_\text{M} \rangle$ as the uniquely invariant state;
\item Spectrum condition: the spectrum of the generators $\hat{P}^\mu$ of the translation are such that $P^0 \geq 0$ (i.e., the energy is nonnegative) and $(P^0)^2 \geq |\vec{P}|^2$ (i.e., the spectrum  of the energy-momentum is confined to the forward light cone, capturing the notion of luminal and subluminal propagation of physical effect);
\item Weak Additivity: for any region $\mathcal{O} \subseteq \mathcal{M}$, $\mathfrak{A}(\mathcal{M})$ is the smallest algebra containing $\bigcup_{\alpha^\mu \in \mathbb{R}^4} \mathfrak{A}(\mathcal{O}_\alpha)$, where $\mathcal{O}_\alpha$ is the region $\mathcal{O}$ translated by $\alpha^\mu$.
\end{enumerate}

Axioms 3-5 are used to prove the Reeh-Schlieder theorem \cite{Reeh:1961ujh, haag1992local}. The theorem states that the vacuum $|0_\text{M} \rangle$ is cyclic for any local algebra $\mathfrak{A}(\mathcal{O})$, in the sense that for any region $\mathcal{O}$, for any state $| \psi \rangle$ and for any $\epsilon > 0$, there exist an operator $\hat{O} \in \mathfrak{A}(\mathcal{O})$ such that $\parallel \hat{O} |0_\text{M} \rangle - | \psi \rangle \parallel < \epsilon $, where $\parallel \cdot \parallel$ is the norm in the Hilbert space. This means that one can approximate any state of the global Hilbert space with arbitrary precision by applying an element of any local algebra $\mathfrak{A}(\mathcal{O})$ to the vacuum $|0_\text{M} \rangle$. Such an effect is the result of entangled correlations in the vacuum \cite{haag1992local, Redhead1995-REDMAA-2, PhysRevA.58.135}.

By operating in any bounded spacetime region $\mathcal{O}_\text{A}$, one is able to produce any global state $| \psi \rangle$ that may, in principle, differ from $|0_\text{M} \rangle$ in another spacelike separated region $\mathcal{O}_\text{B}$. Even if $\mathcal{O}_\text{A}$ and $\mathcal{O}_\text{B}$ are not causally connected, the restriction of $| \psi \rangle$ in $\mathfrak{A}(\mathcal{O}_\text{B})$ may be different from the restriction of $|0_\text{M} \rangle$ in $\mathfrak{A}(\mathcal{O}_\text{B})$. This result seems to be incompatible with the notion of causality. However, the contradiction is resolved by noticing that the nonlocal effect cannot be used for superluminal signaling. A more detailed discussion will be provided in Sec.~\ref{solving_the_paradox}.

A corollary to the Reeh Schlieder theorem is that the vacuum is a separating state in any local algebra $\mathfrak{A}(\mathcal{O})$, in the sense that for any $\hat{O} \in \mathfrak{A}(\mathcal{O})$, if $\hat{O}$ annihilates the vacuum (i.e., $\hat{O} | 0_\text{M} \rangle = 0$), then $\hat{O}=0$ \cite{Redhead1995-REDMAA-2, RevModPhys.90.045003}. The consequence is that annihilator operators cannot be localized with respect to the AQFT scheme. Hence, there is no local operator that counts particles inside bounded space regions. The number operator $\hat{N}(\mathcal{V})$ defined by \citet{PhysRevD.22.377} is inevitably nonlocal with respect to the AQFT scheme. This also applies to the Newton-Wigner number operator [Eq.~(\ref{N_NewtonWigner})].

\subsection{Apparent violation of causality}\label{solving_the_paradox}

In Sec.~\ref{ReehSchlieder_theorem}, we introduced the apparent violation of causality due to the Reeh Schlieder theorem. To see the problem in a physical scenario, consider two observers, Alice and Bob, which are localized in two spacelike separated regions, $\mathcal{O}_\text{A}$ and $\mathcal{O}_\text{B}$. Alice prepares a state $| \psi \rangle = \hat{O}_\text{A} | 0_\text{M} \rangle$ by means of a local operator $\hat{O}_\text{A} \in \mathfrak{A}(\mathcal{O}_\text{A})$ acting on the vacuum $| 0_\text{M} \rangle$; whereas Bob performs local measurements by means of the observable $\hat{O}_\text{B} \in \mathfrak{A}(\mathcal{O}_\text{B})$. As a consequence of the Reeh Schlieder theorem, we find that there are some cases in which
\begin{equation}\label{local_operator_measurament}
\langle \psi | \hat{O}_\text{B} | \psi \rangle \neq \langle 0_\text{M} | \hat{O}_\text{B} | 0_\text{M} \rangle.
\end{equation}
Equation (\ref{local_operator_measurament}) implies that the preparation of the local state $| \psi \rangle =  \hat{O}_\text{A} | 0_\text{M} \rangle$ in $\mathcal{O}_\text{A}$ can be detected by Bob as a result of measurements of the local observable $\hat{O}_\text{B}$. This seems to be incompatible with the notion of causality since Alice and Bob are spacelike separated.

The problem has been addressed by different authors \cite{Redhead1995-REDMAA-2, CLIFTON20011, VALENTE2014147, RevModPhys.90.045003} and led to the conclusion that the violation of causality is only apparent. The solution is given by noticing that a global change of the state is only due to selective operations that cannot be used for superluminal signaling. This argument will be detailed in the present subsection.

Firstly, notice that Eq.~(\ref{local_operator_measurament}) does not hold if $\hat{O}_\text{A}$ is unitary. Indeed, by using the unitarity condition $\hat{O}_\text{A}^\dagger \hat{O}_\text{A} = 1$ and the microcausal commutation relation $[\hat{O}_\text{A}, \hat{O}_\text{B}] = 0$, we obtain
\begin{equation}\label{local_unitary_operator_measurament_A}
\langle 0_\text{M} | \hat{O}_\text{A}^\dagger \hat{O}_\text{B} \hat{O}_\text{A} | 0_\text{M} \rangle = \langle 0_\text{M} | \hat{O}_\text{A}^\dagger \hat{O}_\text{A} \hat{O}_\text{B} | 0_\text{M} \rangle = \langle 0_\text{M} | \hat{O}_\text{B} | 0_\text{M} \rangle.
\end{equation}
Explicitly, this means that
\begin{equation}\label{KnightLicht_property}
\langle \psi | \hat{O}_\text{B} | \psi \rangle = \langle 0_\text{M} | \hat{O}_\text{B}  | 0_\text{M} \rangle.
\end{equation}

By following \citet{10.1063/1.1703731} and \citet{10.1063/1.1703925}, we say that the state $| \psi \rangle$ satisfies the strictly localization property if it gives the same expectation values as the vacuum for all measurements in the causal complement of $\mathcal{O}_\text{A}$. Equivalently, we say that $| \psi \rangle$ is strictly localized in $\mathcal{O}_\text{A}$ if Eq.~(\ref{KnightLicht_property}) holds for any $\hat{O}_\text{B} \in \mathfrak{A}(\mathcal{O}_\text{B})$ and for any region $\mathcal{O}_\text{B}$ spacelike separated from $\mathcal{O}_\text{A}$. As a result of Eq.~(\ref{local_unitary_operator_measurament_A}), we know that any local unitary operator $\hat{U}_\text{A} \in \mathfrak{A}(\mathcal{O}_\text{A})$ produces a strictly localized state $| \psi \rangle = \hat{U}_\text{A} | 0_\text{M} \rangle$ by acting on the vacuum $| 0_\text{M} \rangle$.

In general, the modification of quantum states due to the interaction with experimental instruments (e.g., emitters) is represented by a unitary evolution $| 0_\text{M} \rangle \mapsto \hat{U}_\text{int} | 0_\text{M} \rangle$. However, one can argue that this is not the only way to prepare local states. For instance, one can use the following procedure: (i) let the device interact with the vacuum $| 0_\text{M} \rangle $ to unitarily prepare the state $ \hat{U}_\text{int} | 0_\text{M} \rangle$; (ii) perform the projective measurement $\hat{P}_i$ over a set of subspaces $\mathcal{H}_i$ of the global Hilbert space; (iii) reject all the states that are not elements of (say) $\mathcal{H}_0$. In this way, the experimenter is sure that the resulting state is an element of $\mathcal{H}_0$. The overall operation is said to be selective due to the experimenter's choice of selecting a subensemble after the measurement.

In this scenario, the preparation of the state in $\mathcal{O}_\text{A}$ affects observations in the spacelike separated region $\mathcal{O}_\text{B}$. To see this, consider a local observable $\hat{O}_\text{B} \in \mathfrak{A}(\mathcal{O}_\text{B})$ and assume that $\hat{U}_\text{int} \in \mathfrak{A}(\mathcal{O}_\text{A})$ and $\hat{P}_0 \in \mathfrak{A}(\mathcal{O}_\text{A})$. The normalized state after the preparation is $| \psi \rangle = \hat{O}_\text{A} | 0_\text{M} \rangle$, with
\begin{equation}\label{psi_local_unitary_nonselective}
\hat{O}_\text{A} = \frac{\hat{P}_0 \hat{U}_\text{int}}{\sqrt{\langle 0_\text{M} | \hat{U}_\text{int}^\dagger \hat{P}_0 \hat{U}_\text{int} | 0_\text{M} \rangle}}
\end{equation}
as a local operator in $\mathcal{O}_\text{A}$. The mean value of $\hat{O}_\text{B}$ is
\begin{equation}\label{local_unitary_nonselective_operator_measurament_A}
\langle \psi | \hat{O}_\text{B} | \psi \rangle = \frac{\langle 0_\text{M} | \hat{U}_\text{int}^\dagger \hat{P}_0 \hat{U}_\text{int} \hat{O}_\text{B}  | 0_\text{M} \rangle}{\langle 0_\text{M} | \hat{U}_\text{int}^\dagger \hat{P}_0 \hat{U}_\text{int} | 0_\text{M} \rangle},
\end{equation}
which is different from $\langle 0_\text{M} | \hat{O}_\text{B} | 0_\text{M} \rangle$. Hence, in this scenario, Eq.~(\ref{local_operator_measurament}) holds and the state $| \psi \rangle$ is not strictly localized in $\mathcal{O}_\text{A}$. Notice that the inequality $ \hat{P}_0 \neq 1$ is crucial for the proof of Eq.~(\ref{local_operator_measurament}).\footnote{The results of Eqs.~(\ref{local_unitary_operator_measurament_A}) and (\ref{local_unitary_nonselective_operator_measurament_A}) can be extended to the case of general quantum operations with local Kraus operators $\hat{K}_i \in \mathfrak{A}(\mathcal{O}_\text{A})$ \cite{kraus1983states, CLIFTON20011, VALENTE2014147}. The statistical operator describing the state after the operation is
\begin{equation}\label{rho_Kraus}
\hat{\rho} = \frac{\sum_i \hat{K}_i |0_\text{M} \rangle \langle 0_\text{M} | \hat{K}_i^\dagger}{\sum_i \langle 0_\text{M} | \hat{K}_i^\dagger \hat{K}_i |0_\text{M} \rangle}.
\end{equation}
The operation is said to be nonselective only when
\begin{equation}\label{Kraus_nonselective}
\sum_i \hat{K}_i^\dagger \hat{K}_i = 1.
\end{equation}
By using the ciclicity of the trace and the commutation relation between $\hat{K}_i \in \mathfrak{A}(\mathcal{O}_\text{A})$ and $\hat{O} \in \mathfrak{A}(\mathcal{O}_\text{B})$, one can prove that
\begin{equation}\label{Tr_rho_A}
\text{Tr} (\hat{\rho} \hat{O}) = \frac{\sum_i \text{Tr} (\hat{K}_i^\dagger \hat{K}_i|0_\text{M} \rangle \langle 0_\text{M} | \hat{O})}{\sum_i \langle 0_\text{M} | \hat{K}_i^\dagger \hat{K}_i |0_\text{M} \rangle}.
\end{equation}
The right hand side of Eq.~(\ref{Tr_rho_A}) is equal to $\text{Tr} (|0_\text{M} \rangle \langle 0_\text{M} | \hat{O})$ if and only if the Kraus operators satisfy Eq.~(\ref{Kraus_nonselective}).}

\begin{figure}
\includegraphics{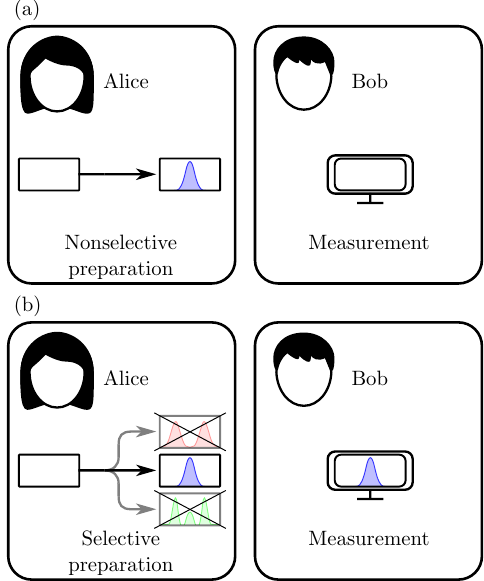}
\caption{Local preparation of states by Alice and local measurement of observables by Bob in two different regions of space. In (a) the preparation of the state is carried out by means of nonselective operations (e.g., via unitary operators) on the vacuum $| 0_\text{M} \rangle$. In this scenario, Bob locally observes the vacuum $| 0_\text{M} \rangle$ as a result of his measurements. In (b) the state is prepared via selective operations (e.g., via projective measurements). Due to the quantum correlations in the vacuum $| 0_\text{M} \rangle$, the outcomes of Bob's measurements are affected by the preparation of the state. Such a nonlocal effect cannot be used for superluminal signaling.} \label{relativistic_Figure}
\end{figure}

We found that only nonselective local operations in $\mathcal{O}_\text{A}$ do not change the vacuum in the causal complement of $\mathcal{O}_\text{A}$. Conversely, the Reeh-Schlieder nonlocal effect and the consequent apparent violation of causality occur when the state is prepared by means of selective operations on $| 0_\text{M} \rangle$. In Fig.~\ref{relativistic_Figure}, we depict this result by considering the experimenters Alice and Bob carrying out local operations in two distinct regions of space.

The defining feature of selective operations is the experimenter's decision to only consider the subspace $\mathcal{H}_0$ and reject all states that give negative results in measuring the effect $\hat{P}_0$. Crucially, the outcomes of the projective measurements are random and only the observer knows when the state has been successfully prepared. This information can only be shared by means of a classical communication. Hence, causality is not violated.

To see that selective operations cannot be used for superluminal signaling, consider again the two experimenters, Alice and Bob, localized in two spacelike separated regions $\mathcal{O}_\text{A}$ and $\mathcal{O}_\text{B}$. Alice performs a selective operation in $\mathcal{O}_\text{A}$ to prepare a state, whereas Bob measures the observable $\hat{O} \in \mathfrak{A}(\mathcal{O}_\text{B})$ in $\mathcal{O}_\text{B}$. In order to prepare the desired state, Alice repeats the operation multiple times and excludes the cases in which the outcome of her selective measurements are unsuccessful, i.e., when the desired state has not been successfully prepared. At this point, Alice is biased, as she knows which operation was successful and which was not. Bob, in principle, is ignorant about the outcome of Alice's operations and, hence, does not know when to perform the measurement with the correct state. He can only acquire this information in two possible ways: (i) by performing Alice's projective measurement to verify if the state is the correct one; however this is only possible if Bob has access to Alice's algebra and, hence, if they are not spacelike separated; (ii) by letting Alice share her information via classical communication, which follows relativistic causality and forbids superluminal signaling.

To connect with the literature, we agree with \citet{CLIFTON20011} and with \citet{VALENTE2014147} who recognized that the problematic operations are selective. However, we give a different argument on why no violation of causality occurs even in the case of selective operations. For \citet{CLIFTON20011}, these operations do not retain full physical meaning, but are partly affected by the purely conceptual operation of selecting subensembles. In other words, the selective component of the operation is regarded as mathematical and nonphysical. This leads to the interpretation of quantum states as partly epistemic entities, where each update of state after a measurement only represents a change of knowledge of the experimenter based on the outcome of the measurement. Conversely, \citet{VALENTE2014147} avoided any interpretation of states, while giving arguments to support the thesis that superluminal signaling of selective operations cannot be controlled. We also showed how these operations cannot be used to instantly send information to another experimenter; however, we used a different argument.

Our approach is inspired by the quantum teleportation technique \cite{PhysRevLett.70.1895}, where a maximally entangled state is used to teleport a quantum state. In that case, no violation of causality occurs because a classical channel has to be employed to transmit information about the outcomes of Alice's measurement. This is in complete analogy with the scenario of the Reeh-Schlieder apparent paradox described here. Hence, we used the argument that the nonlocal correlations due to entanglement is compatible with the prohibitions of superluminal causation. The QFT admits correlations between spacelike separated regions \cite{SUMMERS1985257, Summers:1987fn} while causality is not violated.

The result is also directly connected to the EPR experiment \cite{PhysRev.47.777, CLIFTON20011}, where nonlocal quantum correlations are used to globally change a state via local measurements. The wave function collapse in the EPR scenario cannot be exploited to instantly send information to another experimenter.  Equivalently, in the Reeh-Schlieder scenario, one uses the vacuum correlations to produce a nonlocal effect, which, however, does not lead to superluminal signaling.

\section{Modal scheme}\label{Modal_localization_scheme}

In the familiar formulation of QFT in Minkowski spacetime \cite{Wald:1995yp}, single particle states are represented as positive frequency modes of the Klein-Gordon equation
\begin{equation} \label{Klein_Gordon}
\left[ \eta^{\mu\nu} \partial_\mu \partial_\nu - \left( \frac{mc}{\hbar} \right)^2 \right] \hat{\phi}(t,\vec{x}) = 0.
\end{equation}
Specifically, states with defined momentum $| \vec{k} \rangle$ are represented by the free modes
\begin{equation}\label{free_modes}
f(\vec{k},t,\vec{x}) =  \sqrt{\frac{\hbar}{(2\pi)^3 2 \omega(\vec{k})}} e^{-i\omega(\vec{k})t + i\vec{k} \cdot \vec{x}},
\end{equation}
which are positive frequency solutions of Eq.~(\ref{Klein_Gordon}) satisfying the orthonormality condition
\begin{subequations}\label{KG_scalar_product_orthonormality_f}
\begin{align}
& ( f(\vec{k}), f(\vec{k}') )_\text{KG} = \delta^3(\vec{k}-\vec{k}'), \\
& ( f^*(\vec{k}), f^*(\vec{k}') )_\text{KG} = -\delta^3(\vec{k}-\vec{k}'),\\
& ( f(\vec{k}), f^*(\vec{k}') )_\text{KG} = 0.
\end{align}
\end{subequations}
Here,
\begin{align}\label{KG_scalar_product}
( \phi, \phi' )_\text{KG} = & \frac{i}{\hbar} \int_{\mathbb{R}^3} d^3x \left[ \phi^*(t,\vec{x}) \partial_0  \phi'(t,\vec{x}) \right. \nonumber\\
& \left. -  \phi'(t,\vec{x}) \partial_0 \phi^*(t,\vec{x}) \right]
\end{align}
is the Klein-Gordon product of the couple of modes $\phi(t,\vec{x})$ and $\phi'(t,\vec{x})$. The scalar field $\hat{\phi}(t,\vec{x})$ decomposes into the modes $f(\vec{k},t,\vec{x})$ and $f^*(\vec{k},t,\vec{x})$ as
\begin{equation} \label{free_field}
\hat{\phi}(t,\vec{x}) = \int_{\mathbb{R}^3} d^3 k \left[ f(\vec{k},t,\vec{x}) \hat{a}(\vec{k}) + f^*(\vec{k},t,\vec{x}) \hat{a}^\dagger(\vec{k}) \right],
\end{equation}
where $ \hat{a}(\vec{k})$ is the annihilator of the single particle with momentum $\vec{k}$.

The modal representation of $| \vec{k} \rangle$ can be extended to any Minkowski-Fock state $| \psi \rangle \in \mathcal{H}_\text{M}$ by using the decomposition
\begin{equation}\label{free_state_decomposition}
| \psi \rangle  = \sum_{n=1}^\infty \frac{1}{\sqrt{n!}} \int_{\mathbb{R}^{3n}} d^{3n} \textbf{k}_n \tilde{\psi}_n (\textbf{k}_n) \prod_{l=1}^n \hat{a}^\dagger(\vec{k}_l) | 0_\text{M} \rangle + \tilde{\psi}_0 | 0_\text{M} \rangle.
\end{equation}
Here, $\textbf{k}_n = (\vec{k}_1, \dots, \vec{k}_n)$ is a $3n$ vector collecting $n$ momenta and $\tilde{\psi}_n (\textbf{k}_n)$ is the symmetrized $n$-particles wave function of $| \psi \rangle$ in the momentum representation. The modal representative of $| \psi \rangle$ is defined as
\begin{align} \label{free_wave_function}
\psi_n (t, \textbf{x}_n) = & \left( \frac{2 m c^2}{\hbar^2} \right)^{n/2} \int_{\mathbb{R}^{3n}} d^{3n} \textbf{k}_n \tilde{\psi}_n (\textbf{k}_n) \nonumber \\
& \times \prod_{l=1}^n f(\vec{k}_l, t, \vec{x}_l),
\end{align}
where $\textbf{x}_n = (\vec{x}_1, \dots, \vec{x}_n)$. To not get confused with the notation of Sec.~\ref{NewtonWigner_localization_scheme}, we say that $\psi_\text{NW}(t,\vec{x})$ is a Newton-Wigner wave function and $\psi_n (t, \textbf{x}_n)$ is a modal wave function.

It can be argued that $\psi_n (t, \textbf{x}_n)$ does not entail any genuine notion of localization in QFT.  In particular, one can refer to the superluminal spreading of the modal wave function $\psi_n (t, \textbf{x}_n)$ to claim that the modal scheme is not suited for the description of localized relativistic states. Such an instantaneous spreading can be proven by noticing that $\psi_n (t, \textbf{x}_n)$ is a linear combination of products of positive frequency modes. Hence, if the support of $\psi_n (t, \textbf{x}_n)$ is compact at a fixed time $t$, then its time derivative $\partial_0 \psi_n (t, \textbf{x}_n)$ is not compactly supported at the same time $t$ \cite{Afanasev:1996nm}. Consequently, the modal wave function instantly develops infinite tails.

In QFT, the function $\psi_n (t, \textbf{x}_n)$ cannot be associated to the probability to find the $n$ particles in $\textbf{x}_n = (\vec{x}_1, \dots, \vec{x}_n)$. However, this is not true in the NRQM. In Secs.~\ref{Comparison_with_the_relativistic_theory} and \ref{Convergence_of_the_modal_scheme_to_the_Born_scheme}, we will show that both the AQFT and the modal scheme converge to the same localization scheme when states and observables are restricted to the nonrelativistic regime. This means that the modal wave functions acquire a genuine notion of localization only in the nonrelativistic limit.

In the remaining part of this section, we formulate the modal localization scheme in terms of localized states and observables. By definition, the state $| \psi \rangle$ is said to be localized in a volume $\mathcal{V}$ at time $t$ with respect to the modal scheme if the support of $\psi_n (t, \textbf{x}_n)$ is in $\mathcal{V}^n$, in the sense that $\psi_n (t, \textbf{x}_n) = 0$ when there is an $l \in \{ 1, \dots, n \}$ such that $\vec{x}_l \notin \mathcal{V}$.

We now show that there is a natural definition of localized operators based on the localization of states with respect to the modal scheme. We start by considering the Minkowski-Fock state
\begin{equation}\label{free_state_decomposition_antiparticles}
| \psi \rangle  = \hat{a}_\text{mod}^\dagger[\psi] | 0_\text{M} \rangle,
\end{equation}
with
\begin{equation}\label{c_phi}
\hat{a}_\text{mod}^\dagger[\psi]  = \sum_{n=1}^\infty \frac{1}{\sqrt{n!}} \int_{\mathbb{R}^{3n}} d^{3n} \textbf{k}_n \tilde{\psi}_n (\textbf{k}_n) \prod_{l=1}^n \hat{a}^\dagger(\vec{k}_l) + \tilde{\psi}_0.
\end{equation}

The identity relating $\tilde{\psi}_n (\textbf{k}_n)$ to $\psi_n (t, \textbf{x}_n)$ is Eq.~(\ref{free_wave_function}), which can be inverted by means of a Fourier transform as
\begin{align} \label{free_wave_function_inverse}
\tilde{\psi}_n (\textbf{k}_n) = & \left[ \frac{\hbar}{(2 \pi)^3 m c^2} \right]^{n/2} \int_{\mathbb{R}^{3n}} d^{3n} \textbf{x}_n  \psi_n (0, \textbf{x}_n) \nonumber\\
& \times \prod_{l=1}^n \left[ \sqrt{\omega(\vec{k}_l)}  e^{- i \vec{k}_l \cdot \vec{x}_l} \right].
\end{align}
Equation (\ref{free_wave_function_inverse}) can be plugged in Eq.~(\ref{c_phi}) to obtain
\begin{equation}\label{c_phi_space}
\hat{a}_\text{mod}^\dagger[\psi]  =  \sum_{n=0}^\infty \frac{1}{\sqrt{n!}}   \int_{\mathbb{R}^{3n}} d^{3n} \textbf{x}_n  \psi_n (0, \textbf{x}_n)  \prod_{l=1}^n \hat{a}_\text{mod}^\dagger (\vec{x}_l),
\end{equation}
with
\begin{equation}\label{a_mod_a}
\hat{a}_\text{mod} (\vec{x})  = \int_{\mathbb{R}^3} d^3 k  \sqrt{\frac{\hbar \omega(\vec{k})}{(2 \pi)^3 mc^2}}  e^{i \vec{k} \cdot \vec{x}} \hat{a}(\vec{k}) .
\end{equation}

For each $\vec{x}$ we indicate the algebra generated by the operator $\hat{a}_\text{mod} (\vec{x})$ and its adjoint as $\mathfrak{A}_\text{M}^\text{mod}(\vec{x})$. For extended regions of space $\mathcal{V} \subseteq \mathbb{R}^3$, we define $\mathfrak{A}_\text{M}^\text{mod}(\mathcal{V})$ as the algebra generated by the operators $\hat{a}_\text{mod} (\vec{x})$ with $\vec{x} \in \mathcal{V}$. By using Eq.~(\ref{c_phi_space}) and the definition of localized states with respect to the modal scheme, we find that $\hat{a}_\text{mod}^\dagger[\psi] $ is an element of $ \mathfrak{A}_\text{M}^\text{mod}(\mathcal{V})$, with $\mathcal{V}$ as the region in which the state $| \psi \rangle = \hat{a}_\text{mod}^\dagger[\psi] | 0_\text{M} \rangle$ is localized at $t=0$. This naturally leads to the definition of local operators as elements of $ \mathfrak{A}_\text{M}^\text{mod}(\mathcal{V})$ and the identification of $ \mathfrak{A}_\text{M}^\text{mod}(\mathcal{V})$ as a local algebra with respect to the modal scheme.

Notice that, due to the invertibility of Eq.~(\ref{a_mod_a}), any operator $\hat{O}$ acting on the Minkowski-Fock space $\mathcal{H}_\text{M}$ admits a region $\mathcal{V} \subseteq \mathbb{R}^3$ such that $\hat{O} \in \mathfrak{A}_\text{M}^\text{mod}(\mathcal{V})$. If, in particular, $\mathcal{V} = \mathbb{R}^3$, then the operator is said to be global (i.e., nonlocal) with respect to the modal scheme.

\section{Comparison between localization schemes}\label{Comparison_between_localization_schemes}
\begin{table*}
\begin{center}
\centering
\begin{tabular}{| >{\raggedright\arraybackslash}m{24em} || >{\centering\arraybackslash}m{5.5em} | >{\centering\arraybackslash}m{10em} |  >{\centering\arraybackslash}m{3.5em} |}
\hline
 & Newton-Wi-gner scheme & AQFT scheme & modal scheme\\
\hline
\hline
Relativistic covariance and causality hold & No & Yes & No \\
\hline
The variable $\vec{x}$ is a genuine position coordinate & No & Yes & Yes \\
\hline
Operators in disjoint spatial regions commute & Yes & Yes & No \\
\hline
States in disjoint spatial regions are orthogonal & Yes & No & No \\
\hline
The global Hilbert space factorizes into local Hilbert spaces: $\mathcal{H}_\text{M} = \bigotimes_i \mathcal{H}_\text{M}(\mathcal{V}_i) $ & Yes & Yes & No \\
\hline
The global vacuum $| 0_\text{M} \rangle$ is entangled across the local Hilbert spaces $ \mathcal{H}_\text{M}(\mathcal{V}_i) $ & No & Yes & - \\
\hline
The local Hilbert spaces $ \mathcal{H}_\text{M}(\mathcal{V}_i) $ are Fock spaces with local vacua $| 0_\text{M}(\mathcal{V}_i) \rangle$ & Yes & No & - \\
\hline
The global vacuum factorizes into the local vacua: $| 0_\text{M} \rangle = \bigotimes_i | 0_\text{M}(\mathcal{V}_i) \rangle $ & Yes & - & - \\
\hline
Localized states live in local Hilbert spaces [Eq.~(\ref{NW_localization_decomposition})] & Yes & - & - \\
\hline
The strict localization property [Eq.~(\ref{KnightLicht_property})] at $t=0$ is guaranteed & Yes & Only for local nonse- lective preparations & No \\
\hline
\end{tabular}
\end{center}
\caption{Summary table of the differences between the Newton-Wigner, the AQFT and the modal localization schemes.}\label{Comparison_between_localization_schemes_Table} 
\end{table*}

In the previous sections, we introduced three different localization schemes in QFT. Here, we compare them and we detail the relevant differences. A summary of the discussion can be found in Table \ref{Comparison_between_localization_schemes_Table}.

\subsection{Newton-Wigner and AQFT scheme}\label{Comparison_between_the_NewtonWigner_and_the_AQFT_schemes}

\subsubsection{Fundamental differences}

There are significant conceptual differences between the Newton-Wigner localization and the AQFT localization schemes. The former is based on the orthogonality between states localized in different regions and leads to frame-dependent (i.e., noncovariant) features and superluminal phenomena. Conversely, the AQFT localization scheme is causal and covariant and, hence, it is regarded as fundamental in nature. In particular, the commutativity of fields in spacelike separated regions guarantees the independence of measurements. Furthermore, if a state is localized in one region $\mathcal{V}$ of space at $t=0$, there is no mean by which one can instantaneously send information outside the light cone of $\mathcal{V}$.

In the Newton-Wigner scheme, the variable $\vec{x}$ appears as a result of second-quantizing the position operator $\hat{\vec{x}}_\text{NW} $. Conversely, in the AQFT scheme, the variable $\vec{x}$ is associated the coordinate system representing the underling spacetime and, hence, entails a genuine notion of localization.

Other differences between the Newton-Wigner and the AQFT schemes can be obtained by considering the respective local algebras $\mathfrak{A}_\text{M}^\text{NW}(\mathcal{V})$ and $\mathfrak{A}(\mathcal{O})$. However, notice that $\mathcal{V}$ is a subset of $\mathbb{R}^3$, whereas $\mathcal{O}$ is a region of the spacetime $\mathcal{M}$. Hence, a direct comparison between the two schemes can only be made if we restrict $\mathcal{O}$ to a space region at $t=0$. This is possible due to the dynamical structure of the field $\hat{\phi}(x^\mu)$. Indeed, as a consequence of the Klein-Gordon equation (\ref{Klein_Gordon}) and the hyperbolic nature of the spacetime, any region $\mathcal{O}$ admits a minimally extended Cauchy region $\mathcal{C}_\mathcal{O} \subseteq \mathbb{R}^3$ inside the hypersurface $t=0$ such that the field operators $\hat{\phi}(x^\mu)$ inside $\mathcal{O}$ can be written in terms of field operator $\hat{\phi}(x^\mu)$ and its conjugate momentum field $\hat{\pi}(x^\mu) = - \partial_0 \hat{\phi}(x^\mu)$ inside $\mathcal{C}_\mathcal{O}$. Explicitly, this means that $\mathfrak{A}(\mathcal{O}) \subseteq \mathfrak{A}_\text{M}^\text{AQFT}(\mathcal{C}_\mathcal{O})$, where $\mathfrak{A}_\text{M}^\text{AQFT}(\mathcal{V})$ is the algebra generated by the field operators $\hat{\phi}(0,\vec{x})$ and $\hat{\pi}(0,\vec{x})$ with varying $\vec{x} \in \mathcal{V}$. We also define the local algebra in a space point $\mathfrak{A}_\text{M}^\text{AQFT}(\vec{x})$ as the one generated by the fields $\hat{\phi}(0,\vec{x})$ and $\hat{\pi}(0,\vec{x})$ with fixed $\vec{x}$. The Newton-Wigner and the AQFT localization schemes can now be directly compared by means of the algebras $\mathfrak{A}_\text{M}^\text{NW}(\mathcal{V})$ and $\mathfrak{A}_\text{M}^\text{AQFT}(\mathcal{V})$, or, equivalently, by means of $\mathfrak{A}_\text{M}^\text{NW}(\vec{x})$ and $\mathfrak{A}_\text{M}^\text{AQFT}(\vec{x})$.

By comparing the algebras $\mathfrak{A}_\text{M}^\text{NW}(\vec{x})$ and $\mathfrak{A}_\text{M}^\text{AQFT}(\vec{x})$, one can notice that the two notions of locality are not compatible, in the sense that if an operator is Newton-Wigner localized in $\vec{x}$, it cannot be localized with respect to the AQFT scheme as well. Explicitly, we are saying that $\mathfrak{A}_\text{M}^\text{NW}(\vec{x}) \neq \mathfrak{A}_\text{M}^\text{AQFT}(\vec{x})$.

This can be proved by reminding that operators in $\mathfrak{A}_\text{M}^\text{NW}(\vec{x})$ are generated by the Newton-Wigner annihilation operator $\hat{a}_\text{NW}(\vec{x})$ and its adjoint, whereas the algebra $\mathfrak{A}_\text{M}^\text{AQFT}(\vec{x})$ is generated by the field operators $\hat{\phi}(0,\vec{x})$ and $\hat{\pi}(0,\vec{x}) =  - \partial_0 \hat{\phi}(t,\vec{x}) |_{t=0}$. By using Eqs.~(\ref{free_field}), (\ref{free_modes}) and (\ref{a_NW}), we obtain
\begin{align}\label{a_tilde_phi_Pi}
\hat{a}_\text{NW}(\vec{x}) = & \int_{\mathbb{R}^3} d^3 x' \left[ f_{\hat{\phi}\mapsto \text{NW}}(\vec{x} - \vec{x}') \hat{\phi}(0, \vec{x}') \right. \nonumber\\
& \left. + f_{\hat{\pi}\mapsto \text{NW}}(\vec{x} - \vec{x}') \hat{\pi}(0, \vec{x}')  \right],
\end{align}
with
\begin{subequations}\label{f_phi_Pi} 
\begin{align}
& f_{\hat{\phi}\mapsto \text{NW}}(\vec{x}) = \int_{\mathbb{R}^3} d^3 k \frac{\sqrt{\omega (\vec{k})} e^{i \vec{k} \cdot \vec{x}}}{\sqrt{2 \hbar} (2 \pi)^3}, \\
& f_{\hat{\pi}\mapsto \text{NW}}(\vec{x}) = \int_{\mathbb{R}^3} d^3 k \frac{-i e^{i \vec{k} \cdot \vec{x}}}{\sqrt{2 \hbar \omega (\vec{k})} (2 \pi)^3}.
\end{align}
\end{subequations}
From Eq.~(\ref{f_phi_Pi}), it is possible to notice that $f_{\hat{\phi}\mapsto \text{NW}}(\vec{x}) $ and $f_{\hat{\pi}\mapsto \text{NW}}(\vec{x}) $ are supported in the whole space $\mathbb{R}^3$. Consequently, the right hand side of Eq.~(\ref{a_tilde_phi_Pi}) is nonlocal with respect to the AQFT scheme. This means that $\hat{a}_\text{NW}(\vec{x}) \notin \mathfrak{A}_\text{M}^\text{AQFT}(\vec{x})$ and, hence, $\mathfrak{A}_\text{M}^\text{NW}(\vec{x}) \neq \mathfrak{A}_\text{M}^\text{AQFT}(\vec{x})$.

\subsubsection{Local particle content}

An important difference between the two schemes is given by the notion of the vacuum as locally and globally devoid of quanta \cite{Fleming2000-FLERMN}. The Newton-Wigner operators $\hat{a}_\text{NW}(\vec{x}) $ [Eq.~(\ref{a_NW})] annihilate the vacuum, i.e., $\hat{a}_\text{NW}(\vec{x}) | 0_\text{M} \rangle = 0 $, and can be used to define a local number density operator $\hat{N}(\mathcal{V})$ [Eq.~(\ref{N_NewtonWigner})]. Conversely, the corollary of the Reeh-Schlieder theorem forbids the definition of such an operator in the AQFT localization scheme \cite{Redhead1995-REDMAA-2, RevModPhys.90.045003}. In that case, the vacuum is not locally devoid of quanta, but only globally.

Notice that the Newton-Wigner operators $\hat{a}_\text{NW}(\vec{x}) $ satisfy the canonical commutation identity
\begin{subequations}\label{a_NW_commutation}
\begin{align}
& [\hat{a}_\text{NW}(\vec{x}), \hat{a}_\text{NW}^\dagger(\vec{x}') ] = \delta^3(\vec{x}-\vec{x}'), \\
& [\hat{a}_\text{NW}(\vec{x}), \hat{a}_\text{NW}(\vec{x}') ] = 0.
\end{align}
\end{subequations}
Hence, $\hat{a}_\text{NW}(\vec{x}) $ can be interpreted as a local annihilation operator in the Newton-Wigner scheme. Conversely, due to the corollary of the Reeh-Schlieder theorem, local creation and annihilation operators do not exist in the AQFT scheme \cite{Redhead1995-REDMAA-2, RevModPhys.90.045003}.

The existence of local creation and annihilation operators in the Newton-Wigner scheme ensures that the global Fock space factorized into local Fock spaces $\mathcal{H}_\text{M} = \bigotimes_i \mathcal{H}_\text{M}^\text{NW}(\mathcal{V}_i)$, where $\{ \mathcal{V}_i \}$ is any partition of $\mathbb{R}^3$. The vacuum of each Fock space $\mathcal{H}_\text{M}^\text{NW}(\mathcal{V}_i)$ will be denoted as $| 0_{\text{M}}^\text{NW}(\mathcal{V}_i) \rangle$ and is defined by $\hat{a}_\text{NW}(\vec{x}) | 0_{\text{M}}^\text{NW}(\mathcal{V}_i) \rangle = 0$, for any $\vec{x} \in \mathcal{V}_i$. From the definition of $| 0_{\text{M}}^\text{NW}(\mathcal{V}_i) \rangle$ and the fact that the Minkowski vacuum $| 0_\text{M} \rangle$ is always annihilated by $\hat{a}_\text{NW}(\vec{x})$, we find that $| 0_\text{M} \rangle$ is equal to the product state of the local vacua, i.e., $| 0_{\text{M}} \rangle = \bigotimes_i | 0_{\text{M}}(\mathcal{V}_i) \rangle$, and, hence, it is not entangled across the local Hilbert spaces $\mathcal{H}_\text{M}^\text{NW}(\mathcal{V}_i)$. 

In the AQFT scheme, the global Hilbert space $\mathcal{H}_\text{M}$ can be factorized by means of the local field operators $\hat{\phi}(0,\vec{x})$ and $\hat{\pi}(0,\vec{x})$ and their equal-time commutation relations
\begin{subequations}\label{phi_commutation_Lagrangian}
\begin{align}
& \left[ \hat{\phi}(t,\vec{x}), \hat{\pi}(t,\vec{x}') \right] = i \hbar \delta^3(\vec{x} - \vec{x}'), \label{phi_commutation_Lagrangian_a}\\
&  \left[ \hat{\phi}(t,\vec{x}), \hat{\phi}(t,\vec{x}') \right]  = \left[ \hat{\pi}(t,\vec{x}), \hat{\pi}(t,\vec{x}') \right] = 0,
\end{align}
\end{subequations}
which lead to $\mathcal{H}_\text{M} = \bigotimes_i \mathcal{H}_\text{M}^\text{AQFT}(\mathcal{V}_i)$.\footnote{We remark that the factorization $\mathcal{H}_\text{M} = \bigotimes_i \mathcal{H}_\text{M}^\text{AQFT}(\mathcal{V}_i)$ is not mathematically precise and can be considered valid only in some sort of limit. In particular, in the rigorous context of AQFT, the microcausality condition does not guarantee the factorization of the global algebra into the local algebras $\mathfrak{A} = \bigotimes_i \mathfrak{A}_\text{M}^\text{AQFT}(\mathcal{V}_i)$.

However, a weaker version of $\mathfrak{A} = \bigotimes_i \mathfrak{A}_\text{M}^\text{AQFT}(\mathcal{V}_i)$ can be found in those theories that satisfy the so-called split property \cite{Doplicher1984}. The assumption is that for any couple of spacetime regions $\mathcal{O}$ and $\mathcal{O}' \supset \mathcal{O} $ there is a type I von Neumann algebra $\mathfrak{R}$ such that $\mathfrak{A}(\mathcal{O}) \subset \mathfrak{R} \subset \mathfrak{A}(\mathcal{O}')$. The split property has been proven in a variety of models, including free massive scalar field \cite{Buchholz1974}, Dirac, Maxwell, free massless scalar fields \cite{Horujy1979} and free massive fermion fields \cite{Summers1982}.

A weak notion of independence via tensor product of local Hilbert spaces and algebras is present in quantum field theories with split property \cite{10.1007/978-3-030-38941-3_1}. In particular, for any regions $\mathcal{O}_\text{A}$, $\mathcal{O}'_\text{A}$ and $\mathcal{O}_\text{B}$ such that $\mathcal{O}_\text{A} \subset \mathcal{O}'_\text{A}$ and $\mathcal{O}_\text{B}$ is spacelike separated from $\mathcal{O}'_\text{A}$, the following isomorphism holds
\begin{equation}
\mathfrak{A}(\mathcal{O}_\text{A}) \vee \mathfrak{A}(\mathcal{O}_\text{B}) \cong \mathfrak{A}(\mathcal{O}_\text{A}) \otimes \mathfrak{A}(\mathcal{O}_\text{B}), 
\end{equation}
where the left-hand side is the algebra generated by sums and products of elements in $\mathfrak{A}(\mathcal{O}_\text{A}) $ and $ \mathfrak{A}(\mathcal{O}_\text{B}) $ and the right-hand side is the spatial tensor product of the algebras.

The notion of independence via tensor product is weak because one can consider any $\mathcal{O}_\text{A}'$ arbitrarily close to $\mathcal{O}_\text{A}$, but never equal. In other words, the region $\mathcal{O}_\text{A}'$ ensures that $\mathcal{O}_\text{A}$ and $\mathcal{O}_\text{B}$ do not touch at their border; however, one can consider the limiting case in which the two regions $\mathcal{O}_\text{A}$ and $\mathcal{O}_\text{B}$ are arbitrary close. Hence, the factorization $\mathfrak{A} = \bigotimes_i \mathfrak{A}_\text{M}^\text{AQFT}(\mathcal{V}_i)$ can only by formalized in such a limit.} However, a factorization of $\mathcal{H}_\text{M}$ into local Fock space is not possible, due to the nonexistence of local creators and annihilators. Consequently, the local Hilbert spaces $\mathcal{H}_\text{M}^\text{AQFT}(\mathcal{V}_i)$ cannot be Fock spaces and the global vacuum $| 0_\text{M} \rangle$ cannot factorize into local vacua. More precisely, $| 0_\text{M} \rangle$ does not factorize into any set of local states, since it is entangled across the local Hilbert spaces $\mathfrak{A}_\text{M}^\text{AQFT}(\mathcal{V}_i)$ \cite{haag1992local, Redhead1995-REDMAA-2, PhysRevA.58.135}.

\subsubsection{Independence via tensor product of local Hilbert spaces and algebras}

In quantum physics, the independence of physical phenomena is represented by the factorization of states and observables. In the usual prescription, two distinct laboratories, A and B, are supplied with their own Hilbert space $\mathcal{H}_\text{A}$ and $\mathcal{H}_\text{B}$, the respective experimenters prepare the states $| \psi_\text{A} \rangle \in \mathcal{H}_\text{A}$ and $| \psi_\text{B} \rangle \in \mathcal{H}_\text{B}$ and perform the measurement of the observable $\hat{O}_\text{A}$ and $\hat{O}_\text{B}$. The global Hilbert space, state and observable are the respective tensor product $\mathcal{H}_\text{A} \otimes \mathcal{H}_\text{B}$, $|\psi_\text{A} \rangle \otimes | \psi_\text{B} \rangle$ and $\hat{O}_\text{A} \otimes \hat{O}_\text{B}$.

A similar factorization also occurs in the Newton-Wigner and the AQFT schemes. In particular, the global Hilbert space factorizes into local Hilbert spaces, i.e., $\mathcal{H}_\text{M} = \bigotimes_i \mathcal{H}_\text{M}(\mathcal{V}_i) $. Hence, the two laboratories A and B can be represented by local fields in the respective regions $\mathcal{V}_\text{A}$ and $\mathcal{V}_\text{B}$. Here, we use a unifying notation for both schemes by indicating local Hilbert spaces as $\mathcal{H}_\text{M}(\mathcal{V}_i) $. Depending on the circumstances, if we are referring to the Newton-Wigner scheme, then $\mathcal{H}_\text{M}(\mathcal{V}_i) =  \mathcal{H}_\text{M}^\text{NW}(\mathcal{V})$; conversely, for the AQFT scheme, $\mathcal{H}_\text{M}(\mathcal{V}_i) =  \mathcal{H}_\text{M}^\text{AQFT}(\mathcal{V})$.

The factorization of $\mathcal{H}_\text{M}$ into $\mathcal{H}_\text{M}(\mathcal{V}_\text{A}) \otimes \mathcal{H}_\text{M}(\mathcal{V}_\text{B}) \otimes \dots $ allows experimenters in $\mathcal{V}_\text{A}$ and $\mathcal{V}_\text{B}$ to independently prepare and measure states in their own bounded regions. The fact that the experimenter in $\mathcal{V}_\text{A}$ is able to perform measurements independently from $\mathcal{V}_\text{B}$ is made possible by local operators in $\mathcal{V}_\text{A}$ which act as an identity on $\mathcal{H}(\mathcal{V}_\text{B})$.

We remark that the only preparations in $\mathcal{V}_\text{A}$ that are guaranteed to not affect measurements in $\mathcal{V}_\text{B}$ are nonselective. Notwithstanding the factorization $\mathcal{H}_\text{M} = \mathcal{H}_\text{M}(\mathcal{V}_\text{A}) \otimes \mathcal{H}_\text{M}(\mathcal{V}_\text{B}) \otimes \dots $, selective operations may still lead to nonlocal effects as a consequence of the nonunitary state update. The problem has been discussed in Sec.~\ref{solving_the_paradox} for the case of the AQFT scheme. In particular, we showed that the strict localization property of states is not always satisfied as a result of the Reeh-Schlieder theorem.

Below, we will demonstrate that the Newton-Wigner scheme is not affected by these nonlocal effects. In particular, we will show that in the Newton-Wigner scheme the strict localization property is always satisfied and, hence, local measurements in $\mathcal{V}_\text{B}$ are independent of selective preparations of states in $\mathcal{V}_\text{A}$.

\subsubsection{Intrinsic notion of localization}

Due to the factorization of the global Fock state $\mathcal{H}_\text{M} = \bigotimes_i \mathcal{H}_\text{M}^\text{NW}(\mathcal{V}_i)$ and the global vacuum $| 0_{\text{M}} \rangle = \bigotimes_i | 0_{\text{M}}(\mathcal{V}_i) \rangle$ in the Newton-Wigner scheme, we find that any state that is localized in  $\mathcal{V}$ can be written as
\begin{equation}\label{NW_localization_decomposition}
| \psi \rangle = \hat{O} | 0_{\text{M}}(\mathcal{V}) \rangle \otimes \left[ \bigotimes_i | 0_{\text{M}}(\mathcal{V}_i) \rangle \right],
\end{equation}
where, in this case, $\{ \mathcal{V}_i \}$ is a partition of $\mathbb{R}^3 \setminus \mathcal{V}$ and $\hat{O}$ is an operator acting on $ \mathcal{H}_\text{M}^\text{NW}(\mathcal{V})$. The same factorization does not occur for localized states in the AQFT scheme, because the global vacuum does not factorize in $\mathcal{H}_\text{M} = \bigotimes_i \mathcal{H}_\text{M}^\text{AQFT}(\mathcal{V}_i)$.

Equation (\ref{NW_localization_decomposition}) gives a definition of localized states in terms of a local Hilbert space $ \mathcal{H}_\text{M}^\text{NW}(\mathcal{V})$ as the domain of the state. Intuitively, we say that the state $| \psi \rangle$ lives in $ \mathcal{H}_\text{M}^\text{NW}(\mathcal{V})$, while it appears indistinguishable from the vacuum outside the region $\mathcal{V}$. Such a notion of localization can be compared to the one provided in Sec.~\ref{NewtonWigner_scheme_in_QFT} by means of local operators $\hat{O} \in \mathfrak{A}_\text{M}^\text{NW}(\mathcal{V})$ acting on the vacuum $| 0_\text{M} \rangle$. The physical interpretation was that the local state is the result of local operations occurring in $\mathcal{V}$ over the vacuum background $| 0_\text{M} \rangle$. Conversely, Eq.~(\ref{NW_localization_decomposition}) gives a notion of localization that is independent of the preparation of the state.

The intrinsic notion of localization provided by Eq.~(\ref{NW_localization_decomposition}) is missing in the AQFT scheme, which can therefore only rely on the interpretation of localized states in terms of local preparations over the vacuum $| 0_\text{M} \rangle$. In that case, the local state $| \psi \rangle = \hat{O} | 0_\text{M} \rangle$ with $\hat{O} \in \mathfrak{A}_\text{M}^\text{AQFT}(\mathcal{V})$ cannot be said to live inside the local Hilbert space $ \mathcal{H}_\text{M}^\text{AQFT}(\mathcal{V})$.

\subsubsection{Strict localization property and Alice-Bob scenario}

As a consequence of Eq.~(\ref{NW_localization_decomposition}) the strict localization property is always satisfied in the Newton-Wigner scheme, in the sense that any state $| \psi \rangle $ localized in $\mathcal{V}_\text{A}$ with respect to the Newton-Wigner scheme always appears indistinguishable from the vacuum $| 0_\text{M} \rangle $ in any other separated region $\mathcal{V}_\text{B}$. Explicitly this means that for any $| \psi \rangle = \hat{O}_\text{A} | 0_\text{M} \rangle $, with $\hat{O}_\text{A} \in \mathfrak{A}_\text{M}^\text{NW}(\mathcal{V}_\text{A})$, and for any observable $\hat{O}_\text{B} \in \mathfrak{A}_\text{M}^\text{NW}(\mathcal{V}_\text{B})$, Eq.~(\ref{KnightLicht_property}) holds. The proof comes from the factorization of $| \psi \rangle $ and $\hat{O}_\text{B}$ in $\mathcal{H}_\text{M} =  \mathcal{H}_\text{M}^\text{NW}(\mathcal{V}_\text{A}) \otimes \mathcal{H}_\text{M}^\text{NW}(\mathcal{V}_\text{B}) \otimes \dots$ and from the normalization condition $1 = \langle \psi | \psi \rangle = \langle 0_\text{M}(\mathcal{V}_\text{A}) | \hat{O}_\text{A}^\dagger \hat{O}_\text{A} | 0_\text{M} (\mathcal{V}_\text{A}) \rangle$.

The result may be understood in terms of the Alice-Bob scenario presented in Sec.~\ref{solving_the_paradox} for the AQFT scheme. An experimenter (Alice) prepares a state over the vacuum $| 0_\text{M} \rangle$ by means of local Newton-Wigner operators $\hat{a}_\text{NW}(\vec{x}) \in \mathfrak{A}_\text{M}^\text{NW}(\mathcal{V}_\text{A})$. Another experimenter (Bob) performs measurements in a separated region by means of local Newton-Wigner operators $\hat{a}_\text{NW}(\vec{x}) \in \mathfrak{A}_\text{M}^\text{NW}(\mathcal{V}_\text{B})$. From Eq.~(\ref{KnightLicht_property}), we deduce that the outcomes of Bob's measurements are independent of the preparation of the state by Alice.

Hereafter, we will refer to this scenario as the Newton-Wigner Alice-Bob experiment to not get confused with AQFT Alice-Bob experiment presented in Sec.~\ref{solving_the_paradox}. The discussion of Sec.~\ref{solving_the_paradox} led to the conclusion that not all the states that are localized with respect to the AQFT scheme are also strictly localized, at variance with the Newton-Wigner scheme.

The two Alice-Bob scenarios lead to different results. One may ask which one would be applicable in real experiments. We have already remarked that the AQFT localization scheme is fundamental and entails causal processes. Hence, we may be prone to consider the AQFT Alice-Bob experiment as the most relevant one, while the Newton-Wigner Alice-Bob scenario should not be understood as having a genuine notion of localization. The processes considered in the Newton-Wigner case are physically realizable, in the sense that the state prepared by Alice and the observable used by Bob exist; however, they can hardly be interpreted as genuinely local. If, for instance, Alice uses an emitter to produce the state over the vacuum, the correct way to describe the QFT interaction between the device and the field is by means of local unitary evolution, with the AQFT notion of localization. This would motivate the idea of considering the AQFT Alice-Bob scenario as the one genuinely describing two macroscopic experimenter living in disjoint regions of space.

\subsubsection{Orthogonality condition}

By means of Eq.~(\ref{a_tilde_phi_Pi}) we found that the Newton-Wigner and the AQFT schemes are incompatible. This seems to contradict the idea of generality advocated by \citet{RevModPhys.21.400}: the two authors only considered a minimal set of physically motivated postulates to define the notion of localization in RQM.

At least one of the postulates for the Newton-Wigner localization must have been ignored in the AQFT scheme. The missing assumption is the orthogonality of states in different spatial positions. To see this, consider the states $| \psi_\text{A} \rangle = \hat{\phi}(0,\vec{x}_\text{A}) | 0_\text{M} \rangle \in \mathfrak{A}_\text{M}^\text{AQFT}(\vec{x}_\text{A}) $ and $| \psi_\text{B} \rangle = \hat{\phi}(0,\vec{x}_\text{B}) | 0_\text{M} \rangle  \in \mathfrak{A}_\text{M}^\text{AQFT}(\vec{x}_\text{B})$, which are respectively localized in $\vec{x}_\text{A}$ and $\vec{x}_\text{B}$ according to the AQFT scheme. Assume that the two points are different, $\vec{x}_\text{A} \neq \vec{x}_\text{B}$, and, hence, the states are localized in disjoint regions. By following Newton and Wigner's assumptions, one would expect that $\langle \psi_\text{A} | \psi_\text{B} \rangle = 0$; however, this is not true. The inequality $\langle \psi_\text{A} | \psi_\text{B} \rangle \neq 0$ can be checked by computing the $2$-point correlation function
\begin{equation}
\langle 0_\text{M} | \hat{\phi}(0,\vec{x}_\text{A}) \hat{\phi}(0,\vec{x}_\text{B}) | 0_\text{M} \rangle  = \frac{\hbar}{(2\pi)^3 }  \int_{\mathbb{R}^3} d^3 k \frac{e^{ i\vec{k} \cdot (\vec{x}_\text{A}-\vec{x}_\text{B})}}{2 \omega(\vec{k})},
\end{equation}
which is different form zero.

We recognize that the orthogonality condition is not met by the AQFT localization. Consequently, the probability transition associated to the two spatially separated states $|\langle \psi_\text{A} | \psi_\text{B} \rangle|^2$ is different form zero. The result is apparently paradoxical, as it seems that there is a nonvanishing probability for a local state to be found in another disjoint region \cite{Fleming2000-FLERMN}. The paradox is resolved if we assume that in AQFT the definition of localized states can only be given in terms of local preparations over the vacuum $| 0_\text{M} \rangle$.

At the beginning of Sec.~\ref{AQFT_localization_scheme}, we said that $| \psi \rangle$ is a localized state with respect to the AQFT scheme if it is the result of local operations on $| 0_\text{M} \rangle$. The same definition was also provided for the Newton-Wigner scheme in Sec.~\ref{NewtonWigner_scheme_in_QFT}. Then, we found that an intrinsic notion of localization naturally occurs due to Eq.~(\ref{NW_localization_decomposition}), which provides a definition of localized states as elements of the local algebras $ \mathcal{H}_\text{M}^\text{NW}(\mathcal{V})$. This notion of localization only appears in the Newton-Wigner scheme. Conversely, in the AQFT scheme, Eq.~(\ref{NW_localization_decomposition}) does not hold because local vacuum states do not exist; hence, the definition of localized states can only be provided in terms of local preparations over the vacuum $| 0_\text{M} \rangle$.

The quantity $|\langle \psi_\text{A} | \psi_\text{B} \rangle|^2$ should be interpreted as the probability for a state locally prepared in $\vec{x}_\text{A}$ to turn into a state that can be locally prepared in $\vec{x}_\text{B}$. The fact that $\langle \psi_\text{A} | \psi_\text{B} \rangle$ is different from zero implies that $| \psi_\text{B} \rangle$ can be obtained as an outcome of the projective measurement $| \psi_\text{B} \rangle \langle \psi_\text{B} | $ on $| \psi_\text{A} \rangle$, i.e., $| \psi_\text{B} \rangle \propto | \psi_\text{B} \rangle \langle \psi_\text{B} |\psi_\text{A} \rangle$. This means that $| \psi_\text{B} \rangle$ may be prepared in both of the following ways: either (i) via local operation $\hat{\phi}(0,\vec{x}_\text{B})$ on $| 0_\text{M} \rangle$ in $\vec{x}_\text{B}$ or (ii) via local operation $\hat{\phi}(0,\vec{x}_\text{A})$ on $| 0_\text{M} \rangle$ in $\vec{x}_\text{A} $ followed by the projective measurement $| \psi_\text{B} \rangle \langle \psi_\text{B} | $. The apparent paradox comes from the unexpected compatibility between (i) and (ii) notwithstanding the fact that $\vec{x}_\text{A} $ and $\vec{x}_\text{B} $ are different points. However, notice that the operator $| \psi_\text{B} \rangle \langle \psi_\text{B} | $ is nonlocal, i.e., $| \psi_\text{B} \rangle \langle \psi_\text{B} | \notin \mathfrak{A}_\text{M}^\text{AQFT}(\vec{x}_\text{B})$. Due to the nonlocality of the projective operation, one should not be surprised by the compatibility between (i) and (ii).

\subsection{Newton-Wigner and modal scheme}\label{Comparison_between_NewtonWigner_and_modal_schemes}

In this subsection, we detail the differences between the Newton-Wigner and the modal scheme.

As remarked in Sec.~\ref{Comparison_between_the_NewtonWigner_and_the_AQFT_schemes}, the variable $\vec{x}$ in the Newton-Wigner scheme is not a space coordinate and, hence, it does not entail any genuine notion of position. Conversely, in the modal scheme, $\vec{x}$ appears as a space coordinate for the positive frequency modes $f(\vec{k},t,\vec{x})$ that are solutions of the Klein-Gordon equation (\ref{Klein_Gordon}). The representatives $f(\vec{k},t,\vec{x})$ inherit the fundamental notion of spacetime event $(t,\vec{x})$ from the QFT framework. Hence, in analogy to the AQFT scheme, we say that the variable $\vec{x}$ entails a genuine notion of position.

A feature that both localization schemes share is the acausal spreading of the wave functions, which was discussed in Secs.~\ref{NewtonWigner_localization_scheme} and \ref{Modal_localization_scheme}. In particular, the superluminal effect in the Newton-Wigner is a result of the Hegerfeldt theorem [Sec.~\ref{Hegerfeldt_theorem}], which is a no-go theorem for localization schemes that simultaneously satisfy (i) causality, (ii) positivity of energy and (iii) orthogonality condition for states in disjoint spatial regions. The acausal spreading of the modal wave functions $\psi_n (t, \textbf{x}_n)$, instead, was proved in Sec.~\ref{Modal_localization_scheme} by means of the non-localizability of positive frequency modes for finite intervals of time. The Hegerfeldt theorem cannot be applied in this case because the assumption (iii) is missing.

The lack of assumption (iii) can be proved by considering two single particle states $| \psi_\text{A} \rangle = \hat{a}_\text{mod}^\dagger (\vec{x}_\text{A}) | 0_\text{M} \rangle$ and $| \psi_\text{B} \rangle = \hat{a}_\text{mod}^\dagger (\vec{x}_\text{B}) | 0_\text{M} \rangle$ respectively localized in $\vec{x}_\text{A}$ and $\vec{x}_\text{B}$. By using Eq.~(\ref{a_mod_a}) and the commutation relations (\ref{Minkowski_canonical_commutation}), we obtain
\begin{equation}
\langle \psi_\text{A} | \psi_\text{B} \rangle  = \int_{\mathbb{R}^3} d^3 k  \frac{\hbar \omega(\vec{k})}{mc^2}  \frac{e^{i \vec{k} \cdot (\vec{x}_\text{A} - \vec{x_\text{B}})}}{(2 \pi)^3} , 
\end{equation}
which means that $\langle \psi_\text{A} | \psi_\text{B} \rangle$ different from zero even if $\vec{x}_\text{A} \neq \vec{x}_\text{B}$. At variance with the Newton-Wigner scheme, the modal scheme admits non-orthogonal states that are localized in disjoint spatial regions.

By using again Eq.~(\ref{a_mod_a}) and the commutation relations (\ref{Minkowski_canonical_commutation}) one can also prove that
\begin{subequations}\label{a_mod_commutation}
\begin{align}
& [\hat{a}_\text{mod}(\vec{x}), \hat{a}_\text{mod}^\dagger(\vec{x}') ] = \int_{\mathbb{R}^3} d^3 k  \frac{\hbar \omega(\vec{k})}{ mc^2}  \frac{e^{i \vec{k} \cdot (\vec{x} - \vec{x}')}}{(2 \pi)^3} , \label{a_mod_commutation_a}\\
& [\hat{a}_\text{mod}(\vec{x}), \hat{a}_\text{mod}(\vec{x}') ] = 0.
\end{align}
\end{subequations}
Equation (\ref{a_mod_commutation_a}) implies that the modal operators $\hat{a}_\text{mod}(\vec{x})$ and $\hat{a}_\text{mod}^\dagger(\vec{x})$ cannot be interpreted as local annihilation and creation operators, at variance with the Newton-Wigner operators $\hat{a}_\text{NW}(\vec{x})$ and $\hat{a}_\text{NW}^\dagger(\vec{x})$. It also implies that operators localized in disjoint spatial regions generally do not commute. Explicitly, this means that there are operators $\hat{O}_\text{A} \in \mathfrak{A}_\text{M}^\text{mod}(\vec{x}_\text{A})$ and $\hat{O}_\text{B} \in \mathfrak{A}_\text{M}^\text{mod}(\vec{x}_\text{B})$ such that
\begin{equation}\label{O_A_O_B_neq_0}
[\hat{O}_\text{A}, \hat{O}_\text{B}] \neq 0, 
\end{equation}
even if $\vec{x}_\text{A} \neq \vec{x}_\text{B}$.

As a consequence of Eq.~(\ref{O_A_O_B_neq_0}), the global Hilbert space does not factorize into local Hilbert spaces. This means that the modal localization scheme lacks of the notion of independence via tensor product of local Hilbert spaces. Also, local Fock spaces do not exist and the global vacuum cannot factorize into local vacua, since local Hilbert spaces are nonexistent in the first place. Consequently, the strict localization property is not guaranteed in the modal scheme, at variance with the Newton-Wigner scheme.

All of these differences show that the two localization schemes are incompatible. More generally, it is possible to demonstrate that any operator or state that is localized with respect to one scheme it is not localized with respect to the other. The proof is similar to the one provided in the previous subsection for the Newton-Wigner and the AQFT scheme. Consider the operators $\hat{a}_\text{mod} (\vec{x}) $ and $\hat{a}_\text{NW}(\vec{x})$, which generate the respective local algebras $\mathfrak{A}_\text{M}^\text{mod}(\vec{x})$ and $\mathfrak{A}_\text{M}^\text{NW}(\vec{x})$. Use their definitions [Eqs.~(\ref{a_NW}) and (\ref{a_mod_a})] to compute
\begin{equation}\label{a_mod_a_NW}
\hat{a}_\text{mod} (\vec{x})  =  \int_{\mathbb{R}^3} d^3 x' f_{\text{NW} \mapsto \text{mod}}(\vec{x}-\vec{x}')  \hat{a}_\text{NW}(\vec{x}'),
\end{equation}
with
\begin{equation}\label{f_NW_mod}
f_{\text{NW} \mapsto \text{mod}}(\vec{x}) = \int_{\mathbb{R}^3} d^3 k  \sqrt{\frac{\hbar \omega(\vec{k})}{mc^2}} \frac{e^{i \vec{k} \cdot \vec{x}}}{(2 \pi)^3}.
\end{equation}
Notice that the support of $f_{\text{NW} \mapsto \text{mod}}(\vec{x})$ is $\mathbb{R}^3$, which means that $\hat{a}_\text{mod} (\vec{x}) $ is nonlocal with respect to the Newton-Wigner scheme. This proves that $\mathfrak{A}_\text{M}^\text{mod}(\vec{x}) \neq \mathfrak{A}_\text{M}^\text{NW}(\vec{x})$ with the consequent incompatibility between the two schemes.

\subsection{AQFT and modal scheme}\label{Comparison_between_modal_and_AQFT_schemes}

In Secs.~\ref{Comparison_between_the_NewtonWigner_and_the_AQFT_schemes} and \ref{Comparison_between_NewtonWigner_and_modal_schemes}, we detailed the relevant features of the AQFT and the modal scheme, respectively, and we made a comparison with the Newton-Wigner scheme. In this subsection, instead, we use the results of Secs.~\ref{Comparison_between_the_NewtonWigner_and_the_AQFT_schemes} and \ref{Comparison_between_NewtonWigner_and_modal_schemes} to show the differences between the AQFT and the modal scheme.

The incompatibility between the two schemes can be proved by comparing the respective algebras $\mathfrak{A}_\text{M}^\text{AQFT}(\vec{x})$ and $\mathfrak{A}_\text{M}^\text{mod}(\vec{x})$. By plugging Eq.~(\ref{a_tilde_phi_Pi}) in Eq.~(\ref{a_mod_a_NW}) we obtain
\begin{align}
\hat{a}_\text{mod} (\vec{x}) & = \int_{\mathbb{R}^3} d^3 x' \left[ f_{\hat{\phi}\mapsto \text{mod}}(\vec{x} - \vec{x}') \hat{\phi}(0, \vec{x}') \right. \nonumber\\
& \left. + f_{\hat{\pi}\mapsto \text{mod}}(\vec{x} - \vec{x}') \hat{\pi}(0, \vec{x}')  \right],
\end{align}
with
\begin{subequations}
\begin{align}
& f_{\hat{\phi}\mapsto \text{mod}}(\vec{x}) = \int_{\mathbb{R}^3} d^3 k  \frac{\omega (\vec{k}) e^{i \vec{k} \cdot \vec{x}}}{ (2 \pi)^3 \sqrt{2 m c^2}},\\
 & f_{\hat{\pi}\mapsto \text{mod}}(\vec{x}) = \frac{-i}{\sqrt{2 m c^2}} \delta^3(\vec{x}).
\end{align}
\end{subequations}
The fact that $f_{\hat{\phi}\mapsto \text{mod}}(\vec{x})$ has support in the entire space $\mathbb{R}^3$ implies that the modal operators $\hat{a}_\text{mod} (\vec{x})$ are nonlocal with respect to the AQFT scheme, i.e., $\hat{a}_\text{mod} (\vec{x}) \notin \mathfrak{A}_\text{M}^\text{AQFT}(\vec{x})$, which means that $\mathfrak{A}_\text{M}^\text{mod}(\vec{x}) \neq \mathfrak{A}_\text{M}^\text{AQFT}(\vec{x})$.

As remarked in Secs.~\ref{Comparison_between_the_NewtonWigner_and_the_AQFT_schemes} and \ref{Comparison_between_NewtonWigner_and_modal_schemes}, both the AQFT and the modal scheme are characterized by a genuine position coordinate $\vec{x}$ representing the underling Minkowski spacetime. However, at variance with the AQFT scheme, the modal scheme cannot be considered fundamental in nature. This is due to the acausal effects produced by the superluminal spreading of the wave functions. Also, the microcausality axiom does not hold, as it can be noticed from Eq.~(\ref{O_A_O_B_neq_0}). The non commutativity of operators in disjoint spatial regions does not guarantee the statistical independence of measurements in those regions. For these reasons, the modal scheme does not give a genuine notion of localization.

Due to Eq.~(\ref{O_A_O_B_neq_0}), the strict localization property is not always satisfied, which means that Eq.~(\ref{KnightLicht_property}) does not hold for any $\hat{O}_\text{B} \in \mathfrak{A}_\text{M}^\text{mod}(\mathcal{V}_\text{B})$ and any $| \psi \rangle = \hat{O}_\text{A} | 0_\text{A} \rangle$, with $\hat{O}_\text{A} \in \mathfrak{A}_\text{M}^\text{mod}(\mathcal{V}_\text{A})$ and $\mathcal{V}_\text{A} \cap \mathcal{V}_\text{B} = \varnothing$. This also occurs in the AQFT scheme with $\hat{O}_\text{A} \in \mathfrak{A}_\text{M}^\text{AQFT}(\mathcal{V}_\text{A})$ and $\hat{O}_\text{B} \in \mathfrak{A}_\text{M}^\text{AQFT}(\mathcal{V}_\text{B})$, as a consequence of the Reeh-Schlieder theorem [Sec.~\ref{ReehSchlieder_theorem}]. However, in Sec.~\ref{solving_the_paradox}, we showed that the unitarity of the local operator $\hat{O}_\text{B}$ guarantees the validity of the strict localization property (\ref{KnightLicht_property}) in the AQFT scheme. Crucially, the commutation relation $[\hat{O}_\text{A}, \hat{O}_\text{B}] = 0$ and the definition of unitary operators were used to derive Eq.~(\ref{KnightLicht_property}). In the case of the modal localization scheme, the operators $\hat{O}_\text{A}$ and $\hat{O}_\text{B}$ do not commute, which means that Eq.~(\ref{KnightLicht_property}) is not guaranteed anymore.

\section{Localization in the nonrelativistic regime}\label{Localization_in_NRQM}

In the previous section we detailed three localization schemes for the fully relativistic QFT. Among them, only the AQFT scheme gives a genuine notion of localization. In particular, any local experiment can only be faithfully described in the framework of AQFT. The Newton-Wigner and the modal scheme, instead, appear more as mathematical artifices not suited for a genuine description of local phenomena.

In this section, we consider the nonrelativistic limit of QFT and we show that the three localization scheme converge to each other. Hence, in such a regime, the Newton-Wigner and the modal scheme acquire the genuine notion of localization entailed by the AQFT framework.

\begin{figure}
\includegraphics[width=\columnwidth]{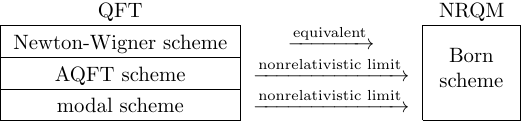}
\caption{Localization schemes in the relativistic (QFT) and the nonrelativistic (NRQM) theory. The Newton-Wigner and the Born scheme are equivalent, whereas the AQFT and the modal scheme converge to the Born scheme in the nonrelativistic limit.} \label{Localization_in_NRQM_Table}
\end{figure}

To obtain this result, we study the NRQM and the notion of localization prescribed by the nonrelativistic theory. We remark that in NRQM the fundamental objects are the first-quantized position and momentum operator, $\hat{x}^i$ and $\hat{k}^i$. The notion of localization in NRQM is based on the definition of $\hat{x}^i$ and on the Born interpretation of quantum mechanics, according to which the modulo square of wave functions gives the probability density to find particles. We demonstrate that such a localization program is equivalent to the Newton-Wigner scheme as they both rely on local creators and annihilators. Then, by following \citet{PhysRevD.90.065032} and \citet{Papageorgiou_2019}, we show that both the AQFT and the modal scheme converge to the Born scheme in the nonrelativistic limit. These results are summarized by Fig.~\ref{Localization_in_NRQM_Table}.

Due to the converge between the Newton-Wigner and the AQFT, we prove that the nonlocal effect described in Sec.~\ref{solving_the_paradox} is suppressed by the nonrelativistic limit. In particular, we show that any state localized in a space region $\mathcal{V}_\text{A}$ is also strictly localized in $\mathcal{V}_\text{A}$, in the sense that it does not affect any measurement conducted in some disjoint region $\mathcal{V}_\text{B}$.

Similarly to Sec.~\ref{Comparison_between_the_NewtonWigner_and_the_AQFT_schemes}, we detail this result by considering an Alice-Bob scenario, in which Alice prepares the state $| \psi \rangle$ in $\mathcal{V}_\text{A}$ and Bob measures $\hat{O}_\text{B}$ in $\mathcal{V}_\text{B}$. At variance with Sec.~\ref{Comparison_between_the_NewtonWigner_and_the_AQFT_schemes}, here, $| \psi \rangle$ and $\hat{O}_\text{B}$ are nonrelativistic and, hence, can be equivalently localized with respect to any scheme. The nonrelativistic Alice-Bob scenario inherits from the AQFT scheme the fundamental notion of localization, in the sense that, regardless of the scheme used to describe the experiment, one always obtains an approximately genuine description of the local phenomena in the two regions $\mathcal{V}_\text{A}$ and $\mathcal{V}_\text{B}$. Also, the strict localization property of the Newton-Wigner Alice-Bob scenario emerges as an independence between the preparation of $| \psi \rangle$ and the measurements of $\hat{O}_\text{B}$.

The section is organized as follows. In Sec.~\ref{Born_localization}, we present the Born scheme, which gives the familiar description of localized states in the NRQM. In Sec.~\ref{Born_localization_is_NewtonWigner_localization}, we show the equivalence between the Newton-Wigner and the Born scheme; whereas, in Secs.~\ref{Comparison_with_the_relativistic_theory} and \ref{Convergence_of_the_modal_scheme_to_the_Born_scheme} we demonstrate the convergence of the AQFT and the modal scheme, respectively, to the Born-Newton-Wigner scheme. In Sec.~\ref{Strict_localization_in_the_nonrelativistic_limit} we use this convergence to show that the Reeh-Schlieder nonlocal effect is suppressed in the nonrelativistic limit and the strict localization property always holds; we detail this result by considering the Alice-Bob scenario in the nonrelativistic regime.

\subsection{Born localization scheme}\label{Born_localization}

In the NRQM, states are localized according to the Born localization principle which assumes that the probability density of finding the system in any space point is the square of the amplitude of its wave function. Hence, particles are localized in the support of their wave functions and they are orthogonal to each other if the localization occurs in disjoint spatial regions. 

In the first-quantized theory, the algebra is generated by the observables position $\hat{x}^i$ and and momentum $\hat{k}^i$, satisfying the canonical commutation relation
\begin{equation}\label{x_p_CCR}
[\hat{x}_i, \hat{k}_j] = i \delta_{i j},
\end{equation}
and by eventual internal degrees of freedom (e.g., spin), which we will ignore for the sake of simplicity. The wave function $\psi(\vec{x}) $ of any state $| \psi \rangle$ can be derived from the eigenstates of $\hat{\vec{x}}$, such that $\psi(\vec{x}) = \langle \vec{x}_\text{B} | \psi \rangle $, where $\hat{x}_i | \vec{x}_\text{B} \rangle = x_i | \vec{x}_\text{B} \rangle$.

Wave functions in the momentum space can be obtained by means of states with defined momentum $| \vec{k} \rangle$, which are defined by $ \hat{k}_i | \vec{k} \rangle = k_i | \vec{k} \rangle$. The identity relating $| \vec{k} \rangle$ to the states with defined position $| \vec{x}_\text{B} \rangle$ is
\begin{equation}\label{x_state_B}
|  \vec{x}_\text{B} \rangle = \int_{\mathbb{R}^3} d^3 k \frac{e^{-i \vec{k} \cdot \vec{x}}}{\sqrt{(2 \pi)^3}} | \vec{k} \rangle,
\end{equation}
which is the Fourier transform of $| \vec{k} \rangle$. One can use Eq.~(\ref{x_state_B}) to switch from the representation of states in the position space to their representation in the momentum space.

From the normalization of $| \psi \rangle$ (i.e., $\langle \psi | \psi \rangle = 1$) and the orthogonality condition $\langle \vec{x}_\text{B} | \vec{x}_\text{B}' \rangle = \delta^3 (\vec{x}-\vec{x}')$, one obtains the familiar result for wave functions
\begin{align}\label{psi_normalization_product}
& \int_{\mathbb{R}^3} d^3 x | \psi(\vec{x}) |^2 = 1, & \langle \psi | \psi' \rangle =  \int_{\mathbb{R}^3} d^3 x \psi^*(\vec{x}) \psi'(\vec{x}) .
\end{align}
Equation (\ref{psi_normalization_product}) captures the idea that $\psi(\vec{x})$ is the probability amplitude of finding the particle in $\vec{x}$, with the consequent interpretation of the support of $\psi(\vec{x})$ as the region of localization for the particle. For any couple of states $| \psi \rangle$ and $| \psi' \rangle$, if $\psi(\vec{x})$ and $\psi'(\vec{x})$ have disjoint support, they are orthogonal to each other.

In the second-quantized theory, the state $| \vec{x}_\text{B} \rangle $ appears as a single particle with defined position. It is defined as
\begin{equation}
| \vec{x}_\text{B} \rangle = \hat{a}_\text{B}^\dagger(\vec{x}) | 0 \rangle,
\end{equation}
with $\hat{a}_\text{B}^\dagger(\vec{x})$ as the creator of the particle in $\vec{x}$ and $| 0 \rangle$ as the vacuum state. All the relevant features of the Born localization scheme in second quantization are inherited from the first-quantized theory. This includes the definition of localized states in terms of compactly supported wave functions and the orthogonality condition for states that are localized in disjoint regions.

By definition, the operators $\hat{a}_\text{B}(\vec{x})$ and $\hat{a}_\text{B}^\dagger(\vec{x})$ satisfy the canonical commutation relation
\begin{align}\label{a_commutation}
& [\hat{a}_\text{B}(\vec{x}), \hat{a}_\text{B}^\dagger(\vec{x}') ] = \delta^3(\vec{x}-\vec{x}'), & [\hat{a}_\text{B}(\vec{x}), \hat{a}_\text{B}(\vec{x}') ] = 0.
\end{align}
As a result of Eq.~(\ref{a_commutation}), the global Fock space factorizes into local Fock spaces $\mathcal{H} = \bigotimes_i \mathcal{H}(\mathcal{V}_i)$ and the global vacuum factorizes into the local vacua $| 0 \rangle = \bigotimes_i | 0 (\mathcal{V}_i) \rangle $. Any state $| \psi \rangle$ localized in $\mathcal{V}$ is equivalently represented by an element of $\mathcal{H}(\mathcal{V})$ such that
\begin{equation}
| \psi \rangle = | \psi (\mathcal{V})  \rangle \otimes \left[ \bigotimes_i | 0 (\mathcal{V}_i) \rangle \right],
\end{equation}
where $| \psi (\mathcal{V}) \rangle$ is the element of $\mathcal{H}(\mathcal{V})$ and $\{ \mathcal{V}_i \}$ is a partition of $\mathbb{R}^3 \setminus \mathcal{V}$. In this sense, we say that the localized state $| \psi \rangle$ lives in the local Fock space $\mathcal{H}(\mathcal{V})$.

\subsection{Equivalence between the Newton-Wigner and the Born scheme}\label{Born_localization_is_NewtonWigner_localization}

In Sec.~\ref{NewtonWigner_localization_scheme}, we presented the Newton-Wigner scheme as an attempt to formalize the notion of localization in RQM and in QFT. Nonrelativistic theories, instead, are described by the Born scheme, which was introduced in Sec.~\ref{Born_localization}.

By comparing the two localization schemes, it is straightforward to see that they are equivalent. In particular, they are both based on the existence of local creation and annihilation operators. All the features found for the Newton-Wigner scheme in Secs.~\ref{NewtonWigner_localization_scheme} and \ref{Comparison_between_the_NewtonWigner_and_the_AQFT_schemes} also apply to the Born localization scheme.

Conceptually, the only difference is given by the regime in which they are defined. The Born scheme was originally introduced in nonrelativistic theories (i.e., NRQM), whereas the Newton-Wigner scheme was conceived in relativistic physics (i.e., RQM and QFT). The original attempt by \citet{RevModPhys.21.400} was precisely to recover the Born interpretation of localized states in the relativistic regime. Consequently, the assumptions postulated by the authors are also met by the Born scheme in NRQM. The results of their work are not only applicable in the relativistic theory but can also be understood in the context of NRQM.

By seeing the NRQM as the nonrelativistic limit of the corresponding field theory, one can embed the Born scheme into the QFT. This leads to a complete equivalence between the Newton-Wigner and the Born scheme in the nonrelativistic regime of quantum fields.

The equivalence is made possible by the fact that both theories are provided with an unifying notion of single particles with defined momentum $| \vec{k} \rangle$. In NRQM, these states are defined as the eigenstates of the first-quantized operator $\hat{\vec{k}}$, whereas in QFT they are associated to a basis of positive frequency solutions of the Klein-Gordon equation (\ref{Klein_Gordon}). The two definitions of $| \vec{k} \rangle$ are unified by the idea that they both represent the same physical object.

By following the Newton-Wigner approach, we define the single particle localized in $\vec{x}$ by means of Eq.~(\ref{x_state}). The same definition applies to the state $| \vec{x}_\text{B} \rangle$ with respect to the Born localization scheme. By comparing Eq.~(\ref{x_state}) with Eq.~(\ref{x_state_B}) we find that $| \vec{x}_\text{B} \rangle = | \vec{x}_\text{NW} \rangle$. This proves the equivalence between the two localization schemes. 

\subsection{Convergence of the AQFT to the Born scheme}\label{Comparison_with_the_relativistic_theory}

In Sec.~\ref{Born_localization}, we remarked that the NRQM is characterized by the Born notion of localization. The NRQM, however, is not regarded as a fundamental theory of physics and it only comes from nonrelativistic approximations of the QFT. Hence, one expects that the Born scheme actually emerges as the nonrelativistic limit of a more fundamental notion of localization properly defined in the QFT.

In Sec.~\ref{Born_localization_is_NewtonWigner_localization}, we showed that the Born localization scheme is equivalent to the Newton-Wigner scheme. In Sec.~\ref{Comparison_between_localization_schemes}, we remarked that QFT has more than one localization scheme and that the Newton-Wigner scheme is in conflict with the notion of localization in AQFT. Between the two schemes, the latter is more fundamental than the former for the following reasons: (i) The genuine notion of position in QFT is given by the Minkowski spacetime upon which the algebra of fields is constructed, while the Newton-Wigner position operator is by no means associated to spacetime events on a manifold; (ii) The AQFT localization scheme is based on the microcausality of fields which forbids violation of causality and superluminal signaling, at variance with the Newton-Wigner which is affected by the instantaneous spreading of wave functions.

In summary, the AQFT scheme gives the fundamental notion of localization in QFT, while the Born scheme defines the localization in the NRQM. The convergence between the two schemes is expected in the nonrelativistic limit as a consequence of the equivalence between the NRQM and the QFT in such a limit. In this section, we show that the Newton-Wigner and the AQFT scheme converge in the nonrelativistic regime. Due to the equivalence between the Newton-Wigner and the Born scheme, this also proves the convergence between the Born and the AQFT scheme.

\subsubsection{Classical versus quantum position}

Before showing the convergence between the two localization schemes, we want to discuss conceptual differences that seem to make them incompatible at any limit. We already remarked that, in AQFT scheme, the variable $\vec{x}$ labeling the fields $\hat{\phi}(0,\vec{x})$ and $\hat{\pi}(0,\vec{x})$ are coordinates representing classical events in the Minkowski spacetime; in this sense we say that the notion of localization in AQFT is classical. Conversely, in NRQM, the variable $\vec{x}$ is used as an index for second-quantized operators generated by the first-quantized position observable $\hat{\vec{x}}$; hence, the notion of localization is quantum. This leads to the apparent incompatibility between the two notions of localization. Why is the position quantum in NRQM and classical in QFT?

An answer to this question can be found by comparing the Galilean and the Poincar\'e group which are at the foundation of the nonrelativistic and the relativistic physics. In NRQM, the operators $\hat{k}^i$ and $m \hat{x}^i$ are, respectively, the generators of the translations and the Galilean boosts. Conversely, in relativistic theories, the Poincar\'e group is defined by the generators of translations $\hat{P}^\mu$, rotations $\hat{J}^i$ and Lorentzian boosts $\hat{K}^i$. It has been proven that in the nonrelativistic limit, the Poincar\'e group converges to the centrally-extended Galilean group and that the generator of Lorentzian boosts $\hat{K}^i$ is approximated by the generator of Galilean boosts $m \hat{x}^i$ \cite{4860f44e-649d-341b-9a70-b912b6531bea, 10.1063/1.524962}. Intuitively, this can be seen by noticing that for small momenta $|\vec{k}| \ll m c / \hbar$, Lorentzian boosts effectively act as Galilean boosts by transforming $\vec{k}$ linearly.

The upshot is that the operator $\hat{x}^i$ should not be interpreted as a quantized version of the Minkowski coordinate $x^i$, but as the limit of the Lorentzian boost operator $\hat{K}^i$ divided by the mass $m$. In the centrally-extended Galilean algebra, the operators $\hat{x}^i$ and $\hat{k}^i$ satisfy the canonical commutation relation (\ref{x_p_CCR}), which leads to the correct transformation rule for the position operator under space translation, i.e.,
\begin{equation}\label{x_translation}
\exp(i \vec{a} \cdot \hat{\vec{k}}) \hat{x}^i \exp(- i \vec{a} \cdot \hat{\vec{k}}) = \hat{x}^i + a^i.
\end{equation}
Consequently, the operator $\hat{x}^i$ plays the dual role of position observable and Galilean boost generator.

The interpretation of $\hat{x}^i$ as a position observable is only valid in the nonrelativistic Galilean theory. The lack of fully relativistic nature in $\hat{x}^i$ is noticeable from the noncovariant and acausal features described in Sec.~\ref{NewtonWigner_localization_scheme}. Notwithstanding the correct behavior under spatial translation [Eq.~(\ref{x_translation})], the operator $\hat{x}^i$ does not properly transform under Lorentz boost and, hence, cannot be seen as a representative of the Poincar\'e group.

We now know why the Born-Newton-Wigner operator $\hat{x}^i$ emerges as a position operator in NRQM. However, in the relativistic theory we already had classical Minkowski coordinates $x^\mu$ assuming the role of position variable. Are they still somehow present in the NRQM or do they disappear in the nonrelativistic limit? The question is conceptually relevant, because, contrary to the operator $\hat{x}^i$, the coordinates $x^\mu$ have a fundamental notion of localization.

Clearly, we cannot directly compare the classical variable $x^\mu$ with the quantum operator $\hat{x}^i$. Instead, we need to consider the second-quantized operators of NRQM labeled by $\vec{x}$ and compare them with field operators of QFT in the hypersurface $t=0$. We remark that second-quantized operators in $\vec{x}$ are generated by the Newton-Wigner operators $\hat{a}_\text{NW}(\vec{x})$, whereas field operators in $(0,\vec{x})$ are generated by $\hat{\phi}(0,\vec{x})$ and $\hat{\pi}(0,\vec{x})$. Alternately, one can consider
\begin{equation}\label{alpha_tilde_phi_Pi}
\hat{a}_\text{AQFT}(\vec{x}) = \sqrt{\frac{m c^2}{2 \hbar^2}} \hat{\phi}(0,\vec{x}) - \frac{i}{\sqrt{2 m c^2}} \hat{\pi}(0,\vec{x}),
\end{equation}
and its adjoin as generators of the local algebra $\mathfrak{A}_\text{M}^\text{AQFT}(\vec{x})$. The inverse of Eq.~(\ref{alpha_tilde_phi_Pi}) is
\begin{subequations}\label{alpha_tilde_phi_Pi_inverse}
\begin{align}
& \hat{\phi}(0,\vec{x})  = \frac{\hbar}{\sqrt{2 m c^2}} \left[ \hat{a}_\text{AQFT}(\vec{x}) + \hat{a}_\text{AQFT}^\dagger(\vec{x}) \right], \\
&  \hat{\pi}(0,\vec{x})  =  i \sqrt{\frac{m c^2}{2}} \left[ \hat{a}_\text{AQFT}(\vec{x}) - \hat{a}_\text{AQFT}^\dagger(\vec{x}) \right].
\end{align}
\end{subequations}

A priori, the variable $\vec{x}$ appearing in $\hat{a}_\text{NW}(\vec{x})$ and in $\hat{a}_\text{AQFT}(\vec{x})$ have different meaning. In the case of $\hat{a}_\text{NW}(\vec{x})$, $\vec{x}$ appears as an index resulting from the second quantization prescription; whereas, in $\hat{a}_\text{AQFT}(\vec{x})$, $\vec{x}$ is a genuine coordinate representing a spacetime event. However, it has been proven that in the nonrelativistic limit, the two fields $\hat{a}_\text{AQFT}(\vec{x})$ and $\hat{a}(\vec{x})$ converge \cite{PhysRevD.90.065032, Papageorgiou_2019}. Consequently, we see that the Minkowski coordinate $\vec{x}$ does not disappear in the nonrelativistic limit, but remains as an index for the annihilator field $\hat{a}(\vec{x})$.

\subsubsection{Convergence between Newton-Wigner and AQFT operators}

The convergence between $\hat{a}_\text{AQFT}(\vec{x})$ and $\hat{a}(\vec{x})$ has two consequences. On one hand, we see that the genuine Minkowski coordinate $\vec{x}$ emerges in the NRQM as an index for the annihilator field $\hat{a}(\vec{x})$. On the other hand, we have the proof that the Newton-Wigner and the AQFT localization schemes converge in the nonrelativistic limit. Indeed, any operator that is localized in $\vec{x}$ with respect to the Newton-Wigner scheme can be approximated by operators which are localized in $\vec{x}$ with respect to the AQFT scheme. This means that the fundamental notion of localization owned by the Minkowski coordinate $\vec{x}$ is approximately shared with the Newton-Wigner position operator $\hat{\vec{x}}$.

To show the convergence between $\hat{a}_\text{AQFT}(\vec{x})$ and $\hat{a}(\vec{x})$, different approaches have been considered, including the use of coarse-grained operators \cite{PhysRevD.90.065032} and the restriction of the Hilbert space to a bandlimited subspace \cite{Papageorgiou_2019}. These methods are based on the definition of a minimum resolution in space and a maximum energy scale by means of the Compton wavelength $\lambda_\text{C} = \hbar/mc$. 

\citet{PhysRevD.90.065032} assume that the minimum experimental resolution of nonrelativistic phenomena is described via coarse-graining modeling. In such a regime, coarse-grained operators are assumed to appear indistinguishable from their fine-grained counterparts. The convergence between the Newton-Wigner and the AQFT localization schemes can be realized by noticing that the kernels $f_{\hat{\phi}\mapsto \text{NW}}(\vec{x}) $ and $f_{\hat{\pi}\mapsto \text{NW}}(\vec{x}) $ appearing in Eq.~(\ref{a_tilde_phi_Pi}) decay exponentially as $\exp(- |\vec{x} |/\lambda_\text{C})$ when $\vec{x}$ is outside the minimum spatial resolution, i.e., $|\vec{x} | \gg \lambda_\text{C}$.

Explicitly, the coarse-grained versions of $\hat{a}_\text{AQFT}(\vec{x})$ and $\hat{a}_\text{NW}(\vec{x})$ are defined as
\begin{subequations}
\begin{align}
& \hat{a}_{\text{AQFT}, \vec{j},\Lambda} = \int_{\mathbb{R}^3} d^3 x G_\Lambda (D \vec{j} - \vec{x}) \hat{a}_\text{AQFT}(\vec{x}), \\
& \hat{a}_{\text{NW},\vec{j},\Lambda} = \int_{\mathbb{R}^3} d^3 x G_\Lambda (D \vec{j} - \vec{x}) \hat{a}_\text{NW}(\vec{x}),
\end{align}
\end{subequations}
with $\vec{j} \in \mathbb{Z}^3$ as grid coordinates, $D$ as the spatial separation of the grid points and 
\begin{equation}
G_\Lambda (\vec{x}) = \frac{1}{(2 \pi \Lambda^2 )^{1/4}} \exp \left( - \frac{|\vec{x}|^2}{4 \Lambda^2} \right)
\end{equation}
as the Gaussian smearing function with spatial resolution $\Lambda \ll D$.  The approximation $\hat{a}_{\text{AQFT}, \vec{j},\Lambda}  \approx \hat{a}_{\text{NW},\vec{j},\Lambda} $ for $\Lambda \gg \lambda_\text{C}$ is proven by \citet{PhysRevD.90.065032} and leads to the convergence between the Newton-Winger and the AQFT schemes in the nonrelativistic limit.

At variance with \citet{PhysRevD.90.065032}, the method adopted by \citet{Papageorgiou_2019} is based on the definition of the bandlimited subspace $\mathcal{H}_\text{M}^\Lambda$ as the Fock space of particles with momenta lower than the cutoff $1/\Lambda$. By restricting $\hat{a}_\text{AQFT}(\vec{x})$ and $\hat{a}_\text{NW}(\vec{x})$ to $\mathcal{H}_\text{M}^\Lambda$ with $\Lambda \gg \lambda_\text{C}$, the authors derive the approximation $\left. \hat{a}_\text{NW}(\vec{x}) \right|_{\mathcal{H}_\text{M}^\Lambda} \approx \left. \hat{a}_\text{AQFT}(\vec{x}) \right|_{\mathcal{H}_\text{M}^\Lambda}$ at first order in $\lambda_\text{C}/\Lambda \ll 1$.

The proof is based on computing the Bogoliubov transformation between the operators $\hat{a}_\text{AQFT}(\vec{x})$ and $\hat{a}(\vec{k})$, i.e.,
\begin{align}\label{alpha_tilde_a}
\hat{a}_\text{AQFT}(\vec{x}) = & \int_{\mathbb{R}^3} d^3 k \left[ f_{\hat{a} \mapsto \text{AQFT}}(\vec{x},\vec{k}) \hat{a}(\vec{k}) \right. \nonumber\\
& \left. + f_{\hat{a}^\dagger \mapsto \text{AQFT}}(\vec{x},\vec{k}) \hat{a}^\dagger(\vec{k})  \right],
\end{align}
with
\begin{subequations}\label{f_a_alpha_tilde}
\begin{align}
& f_{\hat{a} \mapsto \text{AQFT}}(\vec{x},\vec{k}) = \frac{e^{i \vec{k} \cdot \vec{x}}}{2 \sqrt{(2 \pi)^3}}  \left[ \sqrt{\frac{m c^2}{ \hbar \omega(\vec{k})}} +  \sqrt{\frac{ \hbar \omega(\vec{k})}{m c^2}}  \right], \\
& f_{\hat{a}^\dagger \mapsto \text{AQFT}}(\vec{x},\vec{k}) = \frac{e^{-i \vec{k} \cdot \vec{x}}}{2 \sqrt{(2 \pi)^3}}  \left[ \sqrt{\frac{m c^2}{ \hbar \omega(\vec{k})}} - \sqrt{\frac{ \hbar \omega(\vec{k})}{m c^2}}  \right].
\end{align}
\end{subequations}
The restriction of Eqs.~(\ref{a_NW}) and (\ref{f_a_alpha_tilde}) to the bandlimited subspace $\mathcal{H}_\text{M}^\Lambda$ is
\begin{subequations}\label{alpha_tilde_a_a_tilde_a_bandlimited}
\begin{align}
\left. \hat{a}_\text{NW}(\vec{x})  \right|_{\mathcal{H}_\text{M}^\Lambda} = & \int_{|\vec{k}|<1/\Lambda} d^3 k   f_{\hat{a} \mapsto \text{NW}}(\vec{x},\vec{k}) \hat{a}(\vec{k}),\label{a_tilde_a_bandlimited}\\
\left. \hat{a}_\text{AQFT}(\vec{x})  \right|_{\mathcal{H}_\text{M}^\Lambda} = & \int_{|\vec{k}|<1/\Lambda} d^3 k   \left[ f_{\hat{a} \mapsto \text{AQFT}}(\vec{x},\vec{k}) \hat{a}(\vec{k})  \right. \nonumber\\
& \left. + f_{\hat{a}^\dagger \mapsto \text{AQFT}}(\vec{x},\vec{k}) \hat{a}^\dagger(\vec{k})  \right],\label{alpha_tilde_a_bandlimited}
\end{align}
\end{subequations}
with
\begin{equation}\label{f_a_NW}
f_{\hat{a} \mapsto \text{NW}}(\vec{x},\vec{k}) = \frac{e^{i \vec{k} \cdot \vec{x}}}{\sqrt{(2\pi)^3}}.
\end{equation}
Notice that
\begin{subequations}\label{Kernels_approximation}
\begin{align}
& f_{\hat{a} \mapsto \text{AQFT}}(\vec{x},\vec{k}) \approx f_{\hat{a} \mapsto \text{NW}}(\vec{x},\vec{k}) \text{ if } |\vec{k}| \ll \lambda_\text{C}^{-1},\\
&  f_{\hat{a}^\dagger \mapsto \text{AQFT}}(\vec{x},\vec{k}) \approx 0 \text{ if } |\vec{k}| \ll \lambda_\text{C}^{-1}.
\end{align}
\end{subequations}
Hence, by expanding Eqs.~(\ref{alpha_tilde_a_a_tilde_a_bandlimited}) to the first order in $\lambda_\text{C}/\Lambda \ll 1$, we obtain $\left. \hat{a}_\text{NW}(\vec{x}) \right|_{\mathcal{H}_\text{M}^\Lambda} \approx \left. \hat{a}_\text{AQFT}(\vec{x}) \right|_{\mathcal{H}_\text{M}^\Lambda}$, which leads to the convergence between the two localization schemes in the nonrelativistic regime.

\citet{Papageorgiou_2019} also provide the expansion of the Bogoliubov transformation (\ref{alpha_tilde_a_a_tilde_a_bandlimited}) up to the second order in $\lambda_\text{C}/\Lambda \ll 1$. This gives corrective terms that spoil the nonlocality of $\hat{a}_\text{NW}(\vec{x})$ with respect to the AQFT scheme at the first nontrivial order.

\subsection{Convergence of the modal to the Born scheme}\label{Convergence_of_the_modal_scheme_to_the_Born_scheme}

In this subsection, we show the convergence between the modal and the Newton-Wigner scheme in the nonrelativistic limit. Due to the equivalence between the Newton-Wigner and Born scheme [Sec.~\ref{Born_localization_is_NewtonWigner_localization}], we implicitly show the convergence between the modal and the Born scheme.

We follow the strategy of \citet{Papageorgiou_2019} that we already used in Sec.~\ref{Comparison_with_the_relativistic_theory} for the case of the AQFT schemes. Firstly, we restrict Eq.~(\ref{a_mod_a}) to the bandlimited subspace $\mathcal{H}_\text{M}^\Lambda$ to obtain
\begin{align}\label{a_mod_a_Lambda}
\left. \hat{a}_\text{mod} (\vec{x})  \right|_{\mathcal{H}_\text{M}^\Lambda}  =  & \int_{|\vec{k}|<1/\Lambda} d^3 k  \left[ f_{\hat{a} \mapsto \text{AQFT}}(\vec{x},\vec{k}) \right. \nonumber\\
& \left.  - f_{\hat{a}^\dagger \mapsto \text{AQFT}}^*(\vec{x},\vec{k}) \right] \hat{a}(\vec{k}),
\end{align}
where $f_{\hat{a} \mapsto \text{AQFT}}(\vec{x},\vec{k})$ and $f_{\hat{a}^\dagger \mapsto \text{AQFT}}(\vec{x},\vec{k})$ are defined in Eq.~(\ref{f_a_alpha_tilde}). Then, we use Eqs.~(\ref{a_tilde_a_bandlimited}), (\ref{Kernels_approximation}) and (\ref{a_mod_a_Lambda}) to derive the approximation $\left. \hat{a}_\text{NW}(\vec{x}) \right|_{\mathcal{H}_\text{M}^\Lambda} \approx \left. \hat{a}_\text{mod}(\vec{x}) \right|_{\mathcal{H}_\text{M}^\Lambda}$ when $\lambda_\text{C}/\Lambda \ll 1$. This implies that any element of $\mathfrak{A}_\text{M}^\text{mod}(\vec{x})$ can be approximated to an element of $\mathfrak{A}_\text{M}^\text{NW}(\vec{x})$ and that the two localization schemes converge. Due to the equivalence between the Newton-Wigner and Born scheme, we also proved the convergence between the modal and the Born scheme.

We remark that the Born scheme converges to the AQFT scheme as well [Sec.~\ref{Comparison_with_the_relativistic_theory}]. Hence, in this subsection, we have also indirectly proven the convergence between the modal and the AQFT scheme. To have a direct proof, compare Eq.~(\ref{alpha_tilde_a_bandlimited}) with Eq.~(\ref{a_mod_a_Lambda}) and use Eq.~(\ref{Kernels_approximation}) to show that $\left. \hat{a}_\text{AQFT}(\vec{x}) \right|_{\mathcal{H}_\text{M}^\Lambda} \approx \left. \hat{a}_\text{mod}(\vec{x}) \right|_{\mathcal{H}_\text{M}^\Lambda}$. As a consequence of this convergence, we find that the modal scheme acquires a genuine notion of localization in the nonrelativistic regime.

\subsection{The strict localization property in the nonrelativistic limit}\label{Strict_localization_in_the_nonrelativistic_limit}

In Sec.~\ref{solving_the_paradox} we showed that, as a consequence of the Reeh-Schlieder theorem, the AQFT scheme does not always satisfy the strict localization property. This means that the outcome of experiments in any space region $\mathcal{V}_\text{B}$ may depend on the preparation of states in on other disjoint region $\mathcal{V}_\text{A}$. 

At variance with the AQFT scheme, the Newton-Wigner scheme always satisfies the strict localization property [Sec.~\ref{Comparison_between_the_NewtonWigner_and_the_AQFT_schemes}]. However, real life experiments can only be faithfully represented by the AQFT scheme, which is the only one providing a genuine notion of localization. Hence, the strict localization property satisfied by the Newton-Wigner scheme does not generally occur in genuinely local experiments.

The incompatibility between the two schemes disappears in the nonrelativistic limit [Sec.~\ref{Comparison_with_the_relativistic_theory}]. In such a regime, the Newton-Wigner scheme acquires a genuine notion of localization from the AQFT and local experiments are expected to satisfy the strict localization property.

In this subsection, we will show that the nonlocal effects predicted by the AQFT scheme do not occur in the nonrelativistic limit of QFT. Intuitively, the result can be deduced from noticing that Reeh-Schlieder quantum correlations of the vacuum are exponentially suppressed in nonrelativistic scales \cite{SUMMERS1985257}. However, a more detailed proof can be given by using the results of Sec.~\ref{Comparison_with_the_relativistic_theory}. 

The local algebra $\mathfrak{A}_\text{M}^\text{AQFT}(\vec{x})$ is generated by the local fields $\hat{a}_\text{AQFT}(\vec{x})$, which are indistinguishable from the Newton-Wigner fields $\hat{a}_\text{NW}(\vec{x})$ in the nonrelativistic limit. Explicitly, this means that any operator $\hat{O} \in \mathfrak{A}_\text{M}^\text{AQFT}(\vec{x})$ generated by momentum operators $\hat{a}(\vec{k})$ satisfying $\lambda_\text{C} |\vec{k}| \ll 1$ can be approximated to the operator $\hat{O}_\text{NW} \in \mathfrak{A}_\text{M}^\text{NW}(\vec{x})$ obtained by replacing $\hat{a}_\text{AQFT}(\vec{x})$ with $\hat{a}_\text{NW}(\vec{x})$.

Hence, any nonrelativistic state $| \psi \rangle = \hat{O}_\text{A} | 0_\text{M} \rangle$ with $\hat{O}_\text{A} \in \mathfrak{A}_\text{M}^\text{AQFT}(\vec{x}_\text{A})$ and any nonrelativistic observable $\hat{O}_\text{B} \in \mathfrak{A}_\text{M}^\text{AQFT}(\vec{x}_\text{B})$ are approximated by some $| \psi_\text{NW} \rangle = \hat{O}_{\text{NW},\text{A}} | 0_\text{M} \rangle$ and $\hat{O}_{\text{NW},\text{B}} \in \mathfrak{A}_\text{M}^\text{NW}(\vec{x}_\text{B})$,  with $\hat{O}_{\text{NW},\text{A}} \in \mathfrak{A}_\text{M}^\text{NW}(\vec{x}_\text{A})$. The state $| \psi_\text{NW} \rangle$ and the operator $\hat{O}_{\text{NW},\text{B}}$ satisfy the strict localization property
\begin{equation}\label{KnightLicht_property_NW}
\langle \psi_\text{NW} | \hat{O}_{\text{NW},\text{B}} | \psi_\text{NW} \rangle = \langle 0_\text{M} | \hat{O}_{\text{NW},\text{B}}  | 0_\text{M} \rangle.
\end{equation}
when $\vec{x}_\text{A} \neq \vec{x}_\text{B}$. This means that $| \psi_\text{NW} \rangle$ gives the same outcome as the vacuum $| 0_\text{M} \rangle$ when measuring $\hat{O}_{\text{NW},\text{B}}$.

In summary, any state $| \psi \rangle$ that is localized in $\vec{x}_\text{A}$ with respect to the AQFT scheme is approximately localized in $\vec{x}_\text{A}$ with respect to the Newton-Wigner scheme and hence it appears indistinguishable from the vacuum in $\vec{x}_\text{B} \neq \vec{x}_\text{A}$. This means that $| \psi \rangle$ is approximately strictly localized.

To give a practical example, consider the two Alice-Bob scenarios described in Sec.~\ref{Comparison_between_the_NewtonWigner_and_the_AQFT_schemes}. Alice is an experimenter that locally prepares a state in the region $\vec{x}_\text{A}$, while Bob performs local measurements in $\vec{x}_\text{B}$. Depending on the localization scheme, the outcome of Bob's measurements may or may not be influenced by the preparation of the state by Alice.

\begin{figure}
\includegraphics{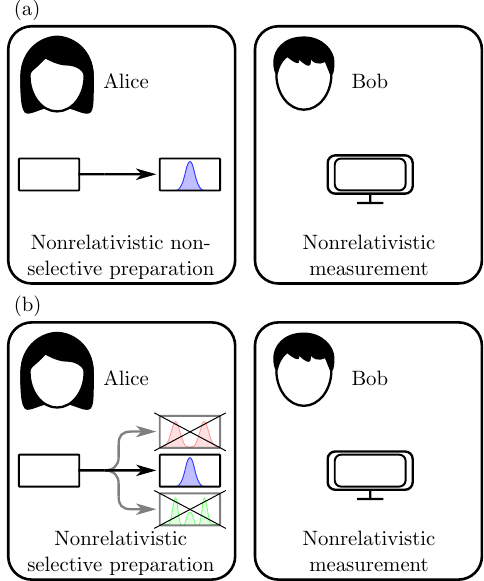}
\caption{Local preparation of nonrelativistic states by Alice and local measurement of nonrelativistic observables by Bob in two different regions of space. The states are prepared via nonselective (a) and selective (b) operations. In both cases, the outcomes of Bob's measurements are not influenced by Alice's local operations. The Reeh-Schlieder nonlocal effect shown in Fig.~\ref{relativistic_Figure} is suppressed by the nonrelativistic limit.} \label{nonrelativistic_Figure}
\end{figure}

In the nonrelativistic limit, the two localization scheme converge. This leads to an equivalence between the two Alice-Bob experiments. In this unifying scenario, the preparation and the measurement in disjoint region appear independent [Fig.~\ref{nonrelativistic_Figure}], in agreement with the Newton-Wigner Alice-Bob experiment. Also, the fundamental notion of localization inherited from the AQFT Alice-Bob scenario guarantees the applicability of the results for genuinely local experiments.

\section{Conclusions}\label{Localization_in_Quantum_Field_Theory_Conclusions}

Different localization schemes have been considered for the QFT in Minkowski spacetime. Among them, only the AQFT framework is able to provide a relativistically consistent notion of localization for states and observables. The Newton-Wigner scheme, instead, is inspired by the nonrelativistic theory and it is based on local creators and annihilators resulting from the definition of a second-quantized position operator. Finally, the modal scheme comes from the modal representation of particles as positive frequency solutions of the Klein-Gordon equation.

Even if the Newton-Wigner and the modal schemes are not suited for the description of relativistic local phenomena, they become indistinguishable from the AQFT scheme in the nonrelativistic limit. Only in the nonrelativistic limit, the familiar description of local states in NRQM by means of wave functions and position operator leads to correct laboratory predictions.

Local preparation of nonrelativistic states never influence nonrelativistic measurements conducted in disjoint space regions. This is at variance with the relativistic regime, where such an independence is violated by selective preparations of states as a consequence of the vacuum correlations contained in the vacuum $| 0_\text{M} \rangle $.

\section*{Acknoweldgement}

We acknowledge financial support from CN1 Quantum
PNRR MUR CN 0000013 HPC and by the HORIZON-EIC-2022-PATHFINDERCHALLENGES-01 HEISINGBERG project 101114978.

\bibliographystyle{apsrmp4-2}
\bibliography{bibliography}

\end{document}